\documentclass[a4paper,11pt]{article}
\usepackage{jcappub}
\usepackage{lineno}
\usepackage{enumitem}
\usepackage{bm}
\usepackage{soul}
\usepackage{comment}
\newcommand{\dd}[0]{\mathrm{d}}
\allowdisplaybreaks

\title{\boldmath 
Dynamics and gravitational radiation of binaries with spin precession and eccentricity in dynamical Chern-Simons gravity}

\author[a,b]{Zhao Li,}
\author[a]{and Wen Zhao}
\affiliation[a]{Department of Astronomy, University of Science and Technology of China, Hefei, Anhui 230026, China; School of Astronomy and Space Sciences, University of Science and Technology of China, Hefei 230026, China}
\affiliation[b]{Department of Physics, Kyoto University, Kyoto 606-8502, Japan}

\emailAdd{lz111301@mail.ustc.edu.cn, wzhao7@ustc.edu.cn}

\abstract{Testing parity symmetry constitutes a critical aspect in gravitational physics. 
As a representative parity-violating theory, dynamical Chern-Simons (dCS) gravity has attracted significant attention in recent gravitational wave (GW) studies. Numerous works have constrained the dCS theory through GW observations using quasi-circular waveform templates.
Since GW parameter estimation depends critically on waveform template accuracy, improved source modeling and waveform construction are essential for tighter constraints on parity-violating gravity. This work explores the dynamics and gravitational radiation from the binary black hole systems with orbital eccentricity and spin precession. By extending the quasi-Keplerian parameterization, we solve the equations of motion including leading-order dCS corrections and precession effects. Furthermore, the conservative sectors of the gravitational and scalar radiation are presented, the corresponding energy and angular momentum loss are calculated, and the orbital decay is also investigated. Notably, because of the non-zero monopole scalar radiation, carrying energy but not angular momentum, the zero-eccentricity orbit is no longer the final stable state of binaries under radiation reaction. This work provides the theoretical foundation for the complete waveform construction in dCS gravity, benefiting the future gravitational parity-symmetry tests.}

\begin{document}
\maketitle
\flushbottom

\section{Introduction}\label{sec:Introduction}

Symmetry plays a central role in physics, providing theories with elegant mathematical structures and profoundly revealing fundamental laws of nature. The covariance of physical laws under parity transformation is one of the crucial discrete symmetries, and parity conservation during scattering and decay processes has long been regarded as a fundamental principle in particle physics. However, in the 1950s, Lee and Yang proposed that parity conservation is broken in weak interactions \cite{Lee1956}, which was subsequently experimentally confirmed by Wu \textit{et al.}, through groundbreaking Cobalt-60 $\beta$-decay experiments in 1957 \cite{Wu1957}. This discovery marked a milestone in modern physics, demonstrating for the first time that parity symmetry could be violated in fundamental interactions.

Inspired by the progress in particle physics, researchers have developed diverse parity-violating modified gravity theories. These theories can be categorized into two classes based on their geometric formulations. The first class is constructed within the Riemannian geometry. For instance, the dynamical Chern-Simons (dCS) gravity introduces a scalar-gravity coupling via the Pontryagin density into the Einstein-Hilbert action \cite{Deser1982,ChernSimons2003,Alexander2009}. The general parity-violating scalar-tensor gravity is proposed to eliminate Ostrogradsky instabilities by involving higher-order derivatives of scalar fields coupled to gravity \cite{Crisostomi2018,WenZhao2020waveform,JinQiao2019,JinQiao2020,JinQiao2023}. The second class is constructed within non-Riemannian geometry. The most prominent examples include teleparallel gravity (TG) \cite{Bahamonde2023} and symmetric teleparallel gravity (STG) \cite{Adak2006}, which describe gravitational interactions through torsion and non-metricity tensors, respectively. The corresponding GR-equivalent formulations are the teleparallel equivalent of general relativity (TEGR) \cite{Maluf2013} and symmetric teleparallel general relativity (STGR) \cite{Nester1999}. Building upon these frameworks, novel parity-violating theories such as Nieh-Yan (NY) gravity \cite{MingzheLi2020NY,MingzheLi2021NY} and parity-violating STG \cite{MingzheLi2022,MingzheLi2022STG} are also proposed, further expanding the landscape of gravitational theories beyond conventional Riemannian descriptions.

This article will mainly focus on dCS gravity. Because the astrophysical models in dCS theory are based on the small-coupling approximation \cite{Yunes2009BH,Yagi2012BH,Yagi2012pn,Yagi2012gw,Yagi2013ns,Loutrel2018,Loutrel2022}, many theoretical predictions apply to the case of $\bar{\alpha}\ll\mathcal{M}_s$, where $\bar{\alpha}$ is the coupling parameter with mass squared dimension and $\mathcal{M}_s$ is the minimum characteristic mass involved in the observed system. To date, the tightest constraint on dCS coupling parameters is given by $\sqrt{\bar{\alpha}}\leqslant8.5\,\mathrm{km}$, achieved from the multimessage observation of neutron stars, including X-ray and gravitational waves (GWs) \cite{Silva2021}. In principle, one can also provide constraints on the dCS theory by observing isolated GW events from binary black hole (BBH) systems. Nair \textit{et al} \cite{Nair2019}, Perkins \textit{et al} \cite{Perkins2021}, and Wang \textit{et al} \cite{HaitianWang2021} presented results for $\sqrt{\bar{\alpha}}\lesssim\mathcal{O}(20-50\,\mathrm{km})$, based on the waveform template of circular-orbit binaries using Bayesian inference. Unfortunately, this result does not meet the small-coupling approximation conditions and is therefore physically meaningless. Through Fisher matrix research, Shi \textit{et al} found that if the circular-orbit waveform template is continued to be used, the third-generation ground-based and space-based GW detectors and arrays cannot effectively constrain the dCS gravity either \cite{ChangfuShi2022}. Beyond the inspiral stage, the ringdown signals from BBH mergers are also a significant channel to test gravitational parity symmetry. Using the parametrized waveform model to measure the events GW150914 and GW200129, with relatively loud ringdown signals, Silva \textit{et al.} placed upper bounds $\sqrt{\bar{\alpha}}\leqslant38.7\,\mathrm{km}$ \cite{Silva2023}. More recently, benefiting from the development of the METRICS framework \cite{Chung2024,Chung2024b}, the quasi-normal mode computation in dCS gravity is beyond the slowly-rotating limit \cite{Chung2025a}. Based on such progress, Chung and Yunes analyzed the events GW150914, GW190521\_074359, and GW200129\_065458, presenting constraints $\sqrt{\bar{\alpha}}\leqslant49\,\mathrm{km}$ \cite{Chung2025}. Whether we can obtain a constraint tighter than $8.5\,\mathrm{km}$ through GW observations is still an open question.

As is well known, GW parameter estimation relies on the accuracy of waveform templates; improvements in wave source modeling and waveform construction are essential to enhance constraints on dCS gravity. This work aims to advance BBH waveform templates in dCS gravity by incorporating orbital eccentricity and spin precession effects. We start from the post-Newtonian (PN) equation of motion (EOM), which originates from the expansion of modified Mathisson-Papapetrou-Dixon (MPD) equations in terms of the typical velocity $v$ (much smaller than the light speed in vacuum) \cite{ZhaoLi2023,Loutrel2018}. As demonstrated in our previous work, dCS gravity modifies the spin-spin coupling and monopole-quadrupole coupling between BBH objects, thereby modifying the evolution of velocity and spin vectors \cite{ZhaoLi2023}. Thus, the EOM consists of the acceleration and precession equations, which are coupled in the spin-precessing binaries, but possess significantly distinguishable characteristic timescales \cite{Chatziioannou2017}. Therefore, one can decouple the acceleration and precession equations at different time scales through multiple-scale analysis, and the binary dynamics is decomposed into orbital motion and Euler rotation of the orbital plane. 

This work mainly focuses on orbital motion and represents the Euler rotation by three unsolved variables, which can be obtained by solving the precession equation under multiple-scale analysis and will be comprehensively discussed in our future research. In the acceleration equation, the dominant dCS correction occurs at the 2PN order, causing precession of the orbital angular momentum (OAM) and the periodic oscillations in the OAM magnitude. To provide sufficient integration constants, we follow the scheme by Gergely \textit{et al.} \cite{Gergely1999,Gergely2000,Gergely2003} and introduce the period-averaged OAM magnitude ($\bar{L}$ in short) in dCS gravity. With the help of conserved energy and $\bar{L}$, we rewrite the acceleration equation and present corresponding extended quasi-Keplerian (QK) parameterized solutions based on the approach by Klein \textit{et al} \cite{Klein2010,Klein2018}. Related to the energy and $\bar{L}$ conservation, one describes the binary orbit by two elements, the semi-major axis and eccentricity. With the QK solution in hand, the scalar and gravitational waveforms, energy loss, OAM loss, and secular evolution of orbital elements can be routinely calculated. Due to the non-zero monopole scalar radiation, which carries energy but not angular momentum, the zero-eccentricity orbit is no longer the final stable state of eccentric processing binaries under radiation reaction, which is significantly distinguishable from that in GR. 

More accurate wave source modeling and waveform calculations can, in principle, improve the signal-to-noise ratio and parameter estimation of GW detection. For example, considering eccentricity is expected to enhance the constraint on Einstein-dilaton-Gauss-Bonnet gravity \cite{Moore2020}. Moreover, Loutrel \textit{et al.} showed that the phase and amplitude of the Fourier dCS waveform, with circular and spin-precessing orbits, differ from those in GR for the same binaries by $\sim10$ radians and $\sim1\%$, respectively for both nearly equal-mass and highly spinning systems, for coupling parameter $\sqrt{\bar{\alpha}}<1\,\mathrm{km}$ \cite{Loutrel2022}. This work establishes a theoretical foundation for constructing a more complete waveform template that includes spin precession and eccentricity, which helps test gravitational parity symmetry based on GW observation. 

This article is organized as follows. We will overview a series of basic acknowledgments on dCS gravity and BBH system in Sec \ref{sec:dCS}. The acceleration equation is written in Sec \ref{sec:frame}. The extended QK solution is presented in Sec \ref{sec:QK}, and the conservative scalar and gravitational waveforms are shown in Sec \ref{sec:GW}. The radiation reaction is investigated in Sec \ref{sec:RR}. The main conclusions are summarized in Sec \ref{sec:Conclusions}. Throughout the paper, we work in geometric units in which $c=G=1$, where $c$ is the light speed in the vacuum and $G$ is the Newtonian gravitational constant.

\section{Dynamical Chern-Simons gravity and binary black hole system}\label{sec:dCS}

\subsection{The action and field equation}
The full action of the dCS theory is \cite{ChernSimons2003,Alexander2009}
\begin{equation}
\label{action}
S=\int\dd^4x\sqrt{-g}\left\{(16\pi)^{-1}R+\frac{\bar{\alpha}}{4}\vartheta R\hat{R}+
\bar{\beta}\left[-\frac{1}{2}(\nabla_{\mu}\vartheta)(\nabla^{\mu}\vartheta)+V(\vartheta)\right]+\mathcal{L}_{m}\right\}.
\end{equation}
The gravity is described by the metric tensor $g_{\mu\nu}$, and $\vartheta$ represents an extra scalar field. The first term in Eq.\,(\ref{action}) gives the Einstein-Hilbert action, where $g$ is the determinant of the metric $g_{\mu\nu}$ and $R$ is the Ricci scalar. The second term represents the coupling between the scalar field and the Pontryagin density, where the Pontryagin density $R\hat{R}$ is defined as
\begin{equation}
R\hat{R}=\frac{1}{2}
\varepsilon^{\rho\sigma\alpha\beta}
R_{\nu\mu\alpha\beta}R^{\mu\nu}_{\ \ \alpha\beta},
\end{equation}
with $\varepsilon^{\rho\sigma\alpha\beta}$ being the Levi-Civit\'{a} tensor, and $R_{\mu\nu\alpha\beta}$ being the Riemann tensor. The third term in Eq.\,(\ref{action}) is the Lagrangian of $\vartheta$, where the scalar potential is usually assumed to be zero. $\mathcal{L}_{m}$ is the Lagrangian density of the matter field. Two coupling parameters, $\bar{\alpha}$ and $\bar{\beta}$, are involved in this theory; the former has a mass-squared dimension, representing the coupling strength, and the latter is dimensionless. 

The variation of the full action (\ref{action}) with respect to the metric $g^{\mu\nu}$ yields the modified field equation \cite{ChernSimons2003,Alexander2009},
\begin{equation}
\label{tensor-equation}
R_{\mu\nu}
-\frac{1}{2}g_{\mu\nu}R
+16\pi\bar{\alpha} C_{\mu\nu}
=8\pi\left[T_{\mu\nu}^{(m)}+T_{\mu\nu}^{(\vartheta)}\right],
\end{equation}
where $R_{\mu\nu}$ is Ricci tensor and $C_{\mu\nu}$ is Cotton tensor defined as
\begin{equation}
\label{C-tensor}
C^{\mu\nu}=
-\varepsilon^{\lambda(\mu|\alpha\beta|}
\left[\nabla_{\alpha}R^{\nu)}_{\ \beta}\right]
(\nabla_{\lambda}\vartheta)
-\hat{R}^{\alpha(\mu|\beta|\nu)}
(\nabla_{\alpha}\nabla_{\beta}\vartheta).
\end{equation}
Note that the Cotton tensor $C_{\mu\nu}$ is traceless, $g^{\mu\nu}C_{\mu\nu}=0$, and satisfies the Bianchi identity, $\nabla^{\mu}C_{\mu\nu}=0$. $T_{\mu\nu}^{(m)}$ and $T_{\mu\nu}^{(\vartheta)}$ denote the energy-momentum tensors of the matter field and the dCS scalar field, i.e.,
\begin{equation}
\label{scalar-EMT}
T^{(m)}_{\mu\nu}=-\frac{2}{\sqrt{-g}}\frac{\delta(\sqrt{-g}\mathcal{L}_{m})}{\delta g^{\mu\nu}},\quad\text{and}\quad
T_{\mu\nu}^{(\vartheta)}=\bar{\beta}
\left[(\nabla_{\mu}\vartheta)(\nabla_{\nu}\vartheta)
-\frac{1}{2}g_{\mu\nu}
(\nabla_{\alpha}\vartheta)(\nabla^{\alpha}\vartheta)\right].
\end{equation}

The scalar field equation can also be derived by variation of the action (\ref{action}) to the scalar field $\vartheta$, which is
\begin{equation}
\label{scalar-equation}
\bar{\beta}\nabla_{\alpha}\nabla^{\alpha}\vartheta
=-\frac{\bar{\alpha}}{4}R\hat{R}.
\end{equation}
We would like to mention here that when the coupling $\bar{\beta}$ is $0$, the full action (\ref{action}) reduces to that of the non-dynamical Chern-Simons gravity. In this case, the scalar field equation (\ref{scalar-equation}) becomes an additional differential constraint, i.e., the {\em Pontryagin constraint} on the space of the allowed solutions, $R\hat{R}=0$. This work will not consider this case but only focus on the dCS gravity, in which the parameter $\bar{\beta}\neq 0$.

This modification to GR leads to a series of parity-violating effects. One of the most well-known predictions is the amplitude birefringence \cite{Alexander2009,ZhaoLi2023,JinQiao2023}, where the amplitude of the left-handed circular polarization mode of GWs increases (or decreases) during the propagation while the amplitude of the right-handed mode decreases (or increases). The similar phenomenons, as well as the velocity birefringence, are investigated in other parity-violating theories, such as general parity-violating scalar-tensor theory \cite{JinQiao2019,JinQiao2020}, NY gravity \cite{MingzheLi2020NY,MingzheLi2021NY,Chatzistavrakidis2022,Bombacigno2023}, Ho{\v{r}}ava-Lifshitz gravity \cite{Takahashi2009,TaoZhu2013,AnzhongWang2013}, parity-violating STG \cite{Conroy2019,MingzheLi2022,MingzheLi2022STG}, spatially covariant gravity \cite{XianGao2014, TaoZhu2022, TaoZhu2023}, and reviewed by Refs.\,\cite{WenZhao2020waveform,JinQiao2023,Jenks2023}. This effect greatly promotes the testing of parity symmetry in the gravitational sector by GW observation \cite{Maria2022,TaoZhu2022,Yunes2010PVtest,Mirshekari2012,WenZhao2020test,QiangWu2022,ChengGong2022,YifanWang2021,YifanWang2022,ZhichaoZhao2022}. 

\subsection{Equation of motion of binary black hole system}

A crucial conclusion in dCS gravity is that there is no modification to the Schwarzschild black hole because of the vanishing Pontryagin density in spherically symmetric spacetime \cite{Grumiller2008BH}. To highlight the parity violation of binary dynamics in dCS, this study concentrate on the binaries including two slow-rotating black holes, whose analytical solution is provided by Refs.\,\cite{Yunes2009BH,Yagi2012BH} up to the quadratic order of dimensionless spin parameter..

Loutrel \textit{et al} have investigated the EOM, the modified MPD equations, of the spinning point masses on the curved background within the framework of effective field theory, in which the Lagrangian is written starting from some required symmetries \cite{Loutrel2019, ZhaoLi2023}. Spinning particles contain three acceleration and three rotation degrees of freedom (DOFs), the former described by the 4-momentum of the particle, and the latter by covariant spin angular momentum (SAM) vectors. For simplification, we call the first equation governing the evolution of 4-momentum the acceleration equation, and the second one, governing the evolution of the covariant SAM vector, the precession equation.

This theoretical framework can be applied to a BBH system, consisting of two spinning black holes, denoted by $A=1, 2$, with mass parameters being $m_{A}$, SAM vectors being $\bm{S}_A$, spatial coordinates being $\bm{r}_A$, and 3-velocity being $\bm{v}_{A}$. Using the PN expansion, we reduce the acceleration and precession equations to the one describing the evolution of 3-velocity $\bm{v}_A$ and spatial SAM $\bm{S}_A$ \cite{ZhaoLi2023}. Up to the 2PN order, where the dominant dCS modification appears, the acceleration equation is written as
\begin{equation}
\label{acceleration-equation}
\bm{a}\equiv\frac{\dd \bm{v}}{\dd t}=\bm{a}_{\mathrm{N}}+\delta\bm{a},
\end{equation}
where
\begin{equation}
\label{acceleration-equation-GR}
\bm{a}_{\mathrm{N}}=-\frac{m}{r^2}\hat{\bm{n}},
\end{equation}
and
\begin{equation}
\label{acceleration-equation-dCS}
\begin{aligned}
\delta\bm{a}&=\zeta\left(-\frac{m}{r^2}\right)\left(\frac{m}{r}\right)^2\Bigg\{-\frac{603}{3584}\sum_{A}\frac{m^2}{m_A^2}\frac{1}{m_A^4}
\left[2(\hat{\bm{n}}\cdot\bm{S}_{A})\bm{S}_{A}
-5(\hat{\bm{n}}\cdot\bm{S}_{A})^2\hat{\bm{n}}
+S_A^2\hat{\bm{n}}\right]\\
&+\frac{75}{256}
\frac{1}{\nu}\frac{1}{m_1^2}\frac{1}{m_{2}^2}
\left[(\bm{S}_1\cdot\bm{S}_2)\hat{\bm{n}}
+(\hat{\bm{n}}\cdot\bm{S}_1)\bm{S}_{2}
+(\hat{\bm{n}}\cdot\bm{S}_2)\bm{S}_1
-5(\hat{\bm{n}}\cdot\bm{S}_1)(\hat{\bm{n}}\cdot\bm{S}_2)
\hat{\bm{n}}\right]\Bigg\}.
\end{aligned}
\end{equation}
In the above three equations, $m\equiv m_1+m_2$ is the total mass of the BBH, $\bm{r}=\bm{r}_1-\bm{r}_2$ is the relative position, and $\bm{v}=\bm{v}_1-\bm{v}_2$ is the relative velocity of objects. The separation is defined as $r=|\bm{r}|$ and the unit vector $\hat{\bm{n}}$ is given by $\bm{r}=r\hat{\bm{n}}$. Furthermore, the symmetric mass ratio is $\nu\equiv m_1m_2/m^2$ and the dimensionless coupling constant is
\begin{equation}
\zeta\equiv16\pi\frac{\bar{\alpha}^2}{\bar{\beta}m^4}.
\end{equation}
It should be mentioned that we only show the Newtonian order and the leading-order dCS correction in this article, while omitting the higher-order PN correction of GW, as these effects have been well-studied in numerous previous works \cite{Blanchet1995, Kidder1995,Faye2006,Blanchet2006,Bohe2013,ThorneHartle1985,Bohe2015,Buonanno2013,Cho2021,Cho2022}. The dCS correction $\delta\bm{a}$ consists of two types of interactions, spin-spin (SS) coupling and monopole-quadrupole (MQ) coupling. The former comes from the interaction between the scalar field generated by two black holes, and the latter comes from the interaction between the spin-induced quadrupole moment and monopole moment. The overall factor $(m/r)^2$ before the curly braces of Eq.\,(\ref{acceleration-equation-GR}) indicates that this correction is at the 2PN order.

In terms of the spatial SAM vector $\bm{S}_A$ and up to the leading-order dCS correction, the precession equation for object $1$ is written as \cite{Loutrel2018,Loutrel2022}
\begin{equation}
\label{precession-equation}
\frac{\dd\bm{S}_1}{\dd t}
=\bm{\Omega}_{12}\times\bm{S}_1,\quad\text{with}\quad
\bm{\Omega}_{12}=\bar{\bm{\Omega}}_{12}+\delta\bm{\Omega}_{12}.
\end{equation}
The GR sector is
\begin{equation}
\label{precession-equation-GR}
\bar{\bm{\Omega}}_{12}
=\frac{1}{r^3}\left(2+\frac{3}{2}\frac{m_2}{m_1}\right)\bm{L}_{\rm N}
+\Big\{\text{Higher PN order}\Big\},
\end{equation}
and the dCS correction is
\begin{equation}
\label{precession-equation-dCS}
\delta\bm{\Omega}_{12}
=\frac{\zeta}{r^3}\left\{-\frac{25}{256}
\frac{1}{\nu^2}
\left[\bm{S}_{2}
-3(\hat{\bm{n}}\cdot\bm{S}_{2})
\hat{\bm{n}}\right]
-\frac{603}{1792}
\frac{m^4}{m_1^4}
\frac{m_{2}}{m_{1}}
(\hat{\bm{n}}\cdot\bm{S}_{1})
\hat{\bm{n}}\right\},
\end{equation}
with $\bm{L}_{\rm N}$ being The Newtonian orbital angular momentum of the system. Comparing with the leading-order effect (\ref{precession-equation-dCS}), the dCS correction is at the relative 0.5PN order, i.e., $\mathcal{O}(\delta\bm{\Omega}_{12}/\bar{\bm{\Omega}}_{12})\\\sim\mathcal{O}(v)$, with $v$ being the magnitude of relative velocity. We also would like to mention that only the leading order of the GR sector is shown in Eq.\,(\ref{precession-equation-GR}), which comes from the spin-orbit coupling. The precession equation for object $2$ is obtained by exchanging the subscripts $1$ and $2$ in Eqs.\,(\ref{precession-equation}, \ref{precession-equation-GR}, \ref{precession-equation-dCS}). The precession equation (\ref{precession-equation}) implies the conservation of spin magnitude, which is proven by, for example, for object $1$, 
\begin{equation}
\label{spin-magnitude-conservation}
\dot{S}_{1}\propto\bm{S}_1\cdot\dot{\bm{S}}_1
=\bm{S}_1\cdot\left(\bm{\Omega}_{12}\times\bm{S}_1\right)
=\bm{\Omega}_{12}\cdot\left(\bm{S}_1\times\bm{S}_1\right)=0.
\end{equation}
A similar conservation for object $2$ is given by exchanging subscripts $1$ and $2$. The conserved spin magnitude plays a crucial role in analytically solving the precession equation \cite{Chatziioannou2017}.

\subsection{Conserved energy and angular momentum}
The conservation of binding energy and orbital angular momentum in Newtonian gravity results in the bound orbit of two bodies being a standard ellipse. In PN theory, gravity is only a slight deviation from Newtonian gravity, which allows us to find a conserved energy that is only a slight deviation from Newtonian binding energy. In dCS theory, the binding energy can always be divided into the Newtonian part and the dCS correction as
\begin{equation}
\label{energy}
E=\mu\varepsilon,\quad\varepsilon
=\varepsilon_{\mathrm{N}}
+\delta\varepsilon,
\end{equation}
where the Newtonian part is given by 
\begin{equation}
\label{energy-Newton}
\varepsilon_{\mathrm{N}}
=\frac{1}{2}v^2-\frac{m}{r}.
\end{equation}
The higher-order PN corrections in GR are not shown in Eq.\,(\ref{energy}) and throughout this work, which are determined by guesswork \cite{Will2014,Faye2006} and can be found in Refs.\,\cite{Kidder1995,Memmesheimer2004}. A similar procedure is also applied to dCS correction, and one can find \cite{ZhaoLi2023}
\begin{equation}
\label{energy-dCS}
\begin{aligned}
\delta\varepsilon&=
\zeta\left(\frac{m}{r}\right)^3
\Bigg\{-\frac{25}{256}
\frac{1}{\nu}\frac{1}{m_1^2}\frac{1}{m_{2}^2}
\left[
(\bm{S}_{1}\cdot\bm{S}_{2})
-3(\hat{\bm{n}}\cdot\bm{S}_{1})
(\hat{\bm{n}}\cdot\bm{S}_{2})\right]\\
&+\frac{201}{3584}\frac{m^2}{m_1^2}\frac{1}{m_1^4}
\left[\bm{S}_{1}^2
-3(\hat{\bm{n}}\cdot\bm{S}_{1})^2\right]
+\frac{201}{3584}\frac{m^2}{m_2^2}\frac{1}{m_2^4}
\left[\bm{S}_{2}^2
-3(\hat{\bm{n}}\cdot\bm{S}_{2})^2\right]\Bigg\},
\end{aligned}
\end{equation}
which also consists of two types of interactions, SS and MQ coupling, like that in the acceleration equation (\ref{acceleration-equation-dCS}). The correction term (\ref{energy-dCS}) is at 2PN order comparing with the Newtonian order, through the estimation $\delta\varepsilon/\varepsilon_{\mathrm{N}}\sim(m/r)^2\sim v^4$.

However, the OAM can no longer be conserved separately due to the spin precession. Alternatively, the total angular momentum (TAM) of the system, defined as the sum of OAM and SAM
\begin{equation}
\label{TAM}
\bm{J}=\bm{L}+\bm{S}_1+\bm{S}_2,
\end{equation}
is still required to be a conserved quantity. By taking the time derivative of Eq.\,(\ref{TAM}) and using the precession equation (\ref{precession-equation}), it can be observed that the OAM remains uncorrected at 2PN and quadratic-spin orders \cite{Will2014,Kidder1995,Bohe2015}. Therefore, the OAM is still written as 
\begin{equation}
\label{OAM}
\bm{L}=\mu\bm{h},\quad\bm{h}\equiv r\left(\hat{\bm{n}}\times\bm{v}\right)
\end{equation}
at the Newtonian order and when omitting the higher PN corrections \cite{Kidder1995}.

\subsection{Gravitational radiation from binaries}
In addition to BBH dynamics, dCS theory modifies gravitational radiation and generates scalar radiation. As discussed in Refs.\,\cite{ZhaoLi2023,Yagi2012pn_e}, the scalar radiation field can be given by linearizing the scalar field equation (\ref{scalar-equation}) and multipole-moment expansion, which is 
\begin{equation}
\label{scalar-radiation}
\vartheta=\frac{5}{8}\frac{\mu}{R}\left(\frac{m}{r}\right)^2\frac{\bar{\alpha}}{\bar{\beta}m^2}\frac{1}{\nu}
\left[(\hat{\mathbf{N}}\cdot\tilde{\bm{\Delta}})
(\hat{\mathbf{N}}\cdot\hat{\bm{n}})
+(\hat{\bm{n}}\cdot\tilde{\bm{\Delta}})\right].
\end{equation}
at the leading order of $\bar{\alpha}$, where $\mu\equiv\nu m$ is the reduced mass, and $\tilde{\bm{\Delta}}$ is defined as
\begin{equation}
\tilde{\bm{\Delta}}\equiv
\frac{m_{2}}{m}\frac{\bm{S}_{1}}{m_1^2}
-\frac{m_{1}}{m}\frac{\bm{S}_{2}}{m_2^2}.
\end{equation}
The unit vector $\hat{\mathbf{N}}$ represents the observational direction, and $R$ is the distance between the observer and the binary center of mass (COM). The term proportional to $\hat{N}^i\hat{N}^j$ in Eq.\,(\ref{scalar-radiation}) is anisotropic scalar quadrupole radiation, which carries both energy and angular momentum. The remaining term in Eq.\,(\ref{scalar-radiation}) is monopole radiation, which does not depend on the observer's orientation and carries only energy, not angular momentum.

By multipole-moment expanding and integrating the linearized gravitational field equation (\ref{tensor-equation}), one determines the dCS correction in gravitational radiation at 2PN order. The complete gravitational waveform in the COM frame is written as
\begin{equation}
\label{tensor-radiation}
\bar{h}_{ij}=\frac{2\mu}{R}\left[\xi^{\mathrm{(N)}}_{ij}+\delta\xi_{ij}\right],
\end{equation}
with the Newtonian order being
\begin{equation}
\label{tensor-radiation-GR}
\xi^{\mathrm{(N)}}_{ij}=2\left(v_iv_j-\frac{m}{r}\hat{n}_i\hat{n}_j\right),
\end{equation}
and the dCS correction is separated into SS and MQ couplings as \cite{Yagi2012pn,Yagi2012pn_e,ZhaoLi2023}
\begin{equation}
\label{tensor-radiation-dCS}
\delta\xi^{ij}=\delta\xi_{\rm(SS)}^{ij}+\delta\xi_{\rm(MQ)}^{ij},
\end{equation}
where
\begin{equation}
\begin{aligned}
\delta\xi_{\rm(SS)}^{ij}&=\frac{75}{128}\frac{\zeta}{\nu}\left(\frac{m}{r}\right)^3\frac{1}{m_1^2m_2^2}
\Big\{\hat{n}^i\hat{n}^j
\left[5(\hat{\bm{n}}\cdot\bm{S}_1)(\hat{\bm{n}}\cdot\bm{S}_2)
-(\bm{S}_1\cdot\bm{S}_2)\right]\\
&\qquad-\left[(\hat{\bm{n}}\cdot\bm{S}_1)(\hat{n}^iS_2^j+\hat{n}^jS_2^j)+(\hat{\bm{n}}\cdot\bm{S}_2)(\hat{n}^iS_1^j+\hat{n}^jS_1^j)\right]\Big\},
\end{aligned}
\end{equation}
\begin{equation}
\begin{aligned}
\delta\xi_{\rm(MQ)}^{ij}=-\frac{603}{1792}\zeta\left(\frac{m}{r}\right)^3
\sum_{A}\frac{m^2}{m_A^2}\frac{1}{m_A^4}
\Big\{\hat{n}^i\hat{n}^j\left[(\hat{\bm{n}}\cdot\bm{S}_A)^2-\bm{S}_A^2\right]
-(\hat{\bm{n}}\cdot\bm{S}_A)(\hat{n}^iS_A^j+\hat{n}^jS_A^j)\Big\}.
\end{aligned}
\end{equation}
Once the specific expressions for $\hat{\bm{n}}$ and $\bm{S}_{A}$ are obtained by solving the acceleration equation (\ref{acceleration-equation}) and precession equation (\ref{precession-equation}), the gravitational and scalar waveform without radiation reaction can be obtained from Eqs.\,(\ref{scalar-radiation},\,\ref{tensor-radiation-dCS}).

As discussed in our previous works \cite{ZhaoLi2023,ZhaoLi2024}, the scalar and gravitational radiations are decoupled at the coalescence timescale, much smaller than the cosmic expansion, resulting in no scalar polarization modes and amplitude birefringence in the gravitational waveform. Therefore, the transverse-traceless (TT) gauge can be safely used in our subsequent calculation.

\section{Frames and acceleration equation}\label{sec:frame}

\subsection{Fundamental, orbital frame, and Euler rotation}

This work considers the BBH system with orbital eccentricity and spin precession, in which the SAM and OAM are no longer conserved separately. As a result, the motion of objects is not on a fixed plane for the distant observer, making it more difficult to completely solve the EOM. However, one has the following observation. 
\begin{itemize}
\item One can always establish a non-inertial frame, in which the precession disappears and the black holes are constrained within a fixed plane. Such a non-inertial frame is called the \emph{orbital frame}. 
\item An inertial frame can be established based on the observer and the binary COM, in which the observer is static. This frame is usually referred to as the \emph{fundamental frame}. The fundamental and orbital frames are related through Euler rotation.
\end{itemize}
For simplification, the axes of the fundamental frame are denoted as $\{X,Y,Z\}$. Using the conservation of TAM, one can conveniently choose the $Z$-axis to be aligned with the TAM. The $X$-axis can be selected arbitrarily and $\bm{e}_Y=\bm{e}_Z\times\bm{e}_X$. The direction of the observer in the fundamental frame is 
\begin{equation}
\hat{\mathbf{N}}=(\sin\iota\cos\omega,\sin\iota\sin\omega,\cos\iota),
\end{equation}
where $\iota$ and $\omega$ are constant, representing the inclination and azimuth angle. 

As for the orbital frame, we denote its axes as $\{x,y,z\}$, and the orbital plane is spanned by the relative position and velocity vectors. When only considering the Newtonian-order approximation, the OAM vector is always perpendicular to such an orbital plane, so the $z$-axis of the orbital frame is naturally set along the direction of the OAM. 

\begin{figure}[ht]
\centering
\includegraphics[width=\textwidth]{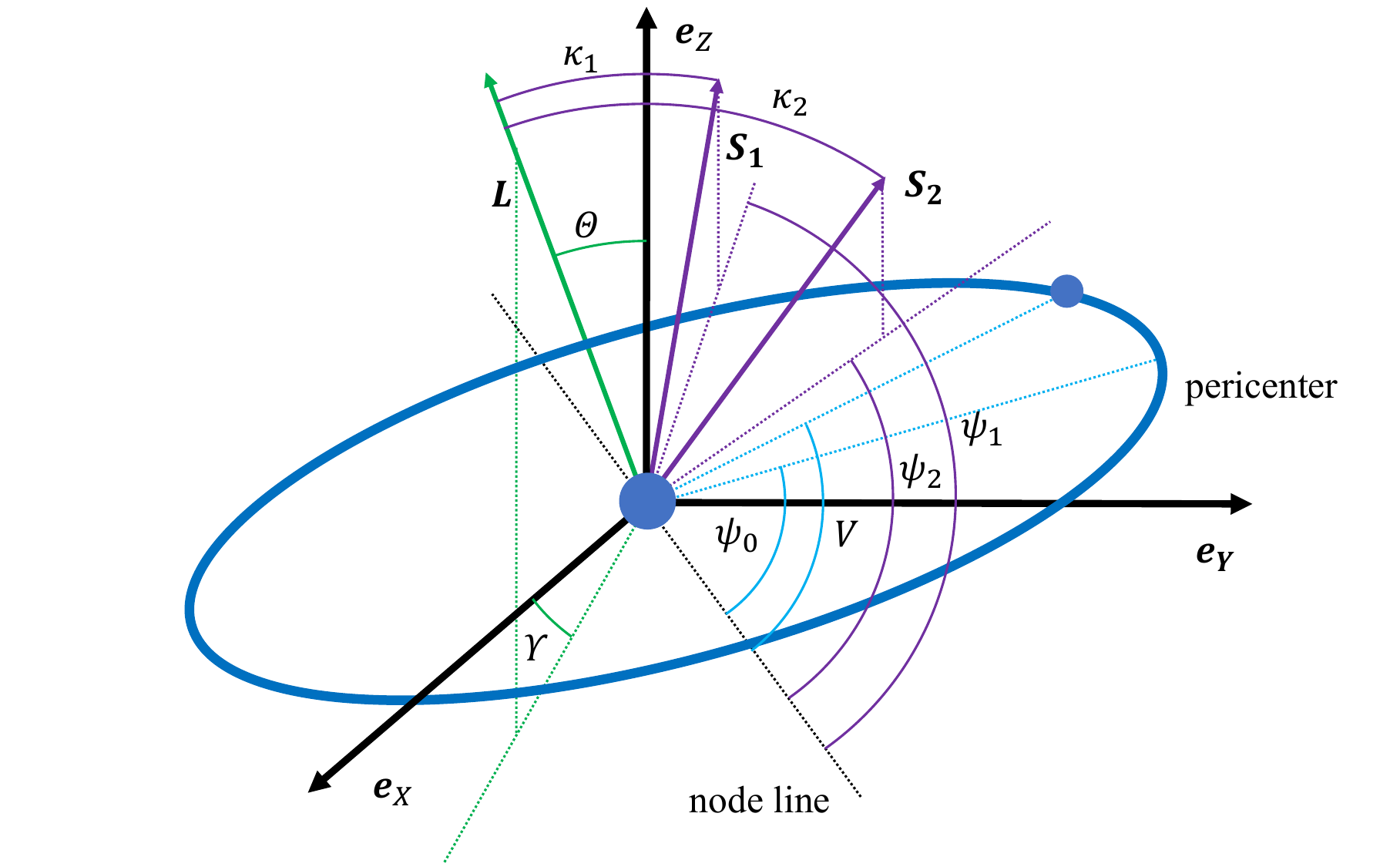}
\caption{The geometry of an eccentric and spin processing BBH system.}
\label{fig:OAM_and_SAM}
\end{figure}

As shown in Fig.\,\ref{fig:OAM_and_SAM}, the \emph{node line} is the intersection between the $x-y$ plane and $X-Y$ plane. The point where the object trajectory passes through the $X-Z$ plane from the negative-$Z$ direction to the positive-$Z$ direction is called the ascending node, and vice versa is called the descending node. The direction from the binary COM to the ascending node is usually defined as the $x$-axis of the orbital frame. Because the node line is perpendicular to both the $z$ and $Z$-axis, therefore $\bm{e}_{x}\propto\bm{e}_Z\times\bm{e}_z$. In addition, it is convenient to define a set of angles:
\begin{itemize}
\item \emph{The inclination} $\Theta$, the angle between the $z$-axis of the orbital frame and the $Z$-axis of the fundamental frame, or the angle between OAM and TAM, equivalently.
\item \emph{The longitude of the ascending node} $\Upsilon$, the azimuth angle of the projected vector of OAM on the $X-Y$ plane.
\item \emph{The longitude of the pericenter} $\psi_0$, the angle between the direction of orbital pericenter and the node line.
\end{itemize}
The transformation between the basis vectors of the fundamental and the orbital frames is a series of Euler rotations. Two of the three Euler angles are the inclination angle $\Theta$ and the right ascension of the ascending node $\Upsilon$. The third one $\alpha$ represents the Lorentz boost between the inertial and non-inertial frames, determined by \cite{Arun2009}
\begin{equation}
\label{alpha-equation}
\dot{\alpha}=-\dot{\Upsilon}\cos\Theta.
\end{equation}
The Euler rotation between fundamental and orbital frames is written as
\begin{equation}
\label{transformation-from-fundamental-to-orbital-frame}
\left(\bm{e}_{x},\,\bm{e}_{y},\,\bm{e}_{z}\right)^{\rm T}
=\bm{\mathcal{R}}\cdot\left(\bm{e}_{X},\,\bm{e}_{Y},\,\bm{e}_{Z}\right)^{\rm T},
\end{equation}
where the rotation matrix is defined as
\begin{equation}
\label{rotation-matrix}
\bm{\mathcal{R}}
=\left(\begin{array}{ccc}
\cos\alpha & \sin\alpha & 0 \\
-\sin\alpha & \cos\alpha & 0 \\
0 & 0 & 1
\end{array}\right)
\left(\begin{array}{ccc}
1 & 0 & 0 \\
0 & \cos \Theta & \sin \Theta \\
0 & -\sin \Theta & \cos \Theta
\end{array}\right)
\left(\begin{array}{ccc}
\cos\Upsilon & \sin\Upsilon & 0 \\
-\sin\Upsilon & \cos\Upsilon & 0 \\
0 & 0 & 1
\end{array}\right).
\end{equation}

\subsection{Rewriting the acceleration equation}
Rather than directly solving the acceleration equation (\ref{acceleration-equation}), we rewrite conserved energy (\ref{energy}) and OAM (\ref{OAM}) as a new set of equations, and then solve them using QK parameterization. In the fundamental frame, the relative position vector and its time derivative are expressed as
\begin{equation}
\bm{r}=r\hat{\bm{n}},\quad
\dot{\bm{r}}=\dot{r}\hat{\bm{n}}+r\dot{\theta}\bm{e}_{\theta}+r\sin\theta\dot{\phi}\bm{e}_{\phi}.
\end{equation}
where $\theta$ and $\phi$ are the spherical coordinates relating to the $\{X,Y,Z\}$-system. With the help of the above quantities, the conserved energy (\ref{energy}) is 
\begin{equation}
\label{energy-fundamental}
\varepsilon=\frac{1}{2}\left[\dot{r}^2+r^2\left(\dot{\theta}^2+\dot{\phi}^2\sin^2\theta\right)\right]-\frac{m}{r}+\delta\varepsilon,
\end{equation}
and the OAM is 
\begin{equation}
\label{OAM-fundamental}
\bm{h}=r^2
\left(\begin{array}{c}
-\dot{\theta}\sin\phi
-\dot{\phi}\sin\theta\cos\theta\cos\phi\\
\dot{\theta}\cos\phi
-\dot{\phi}\sin\theta\cos\theta\sin\phi\\
\dot{\phi}\sin^2\theta
\end{array}\right),
\end{equation}
whose magnitude $h$ and $Z$-component $h\cos\Theta$ are
\begin{equation}
h^2=r^4\left(\dot{\theta}^2
+\dot{\phi}^2\sin^2\theta\right),\quad\text{and}\quad
h\cos\Theta=r^2\dot{\phi}\sin^2\theta,
\end{equation}
respectively. Using the parameters $\varepsilon$, $h^2$, and $h\cos\Theta$, a set of new acceleration equation in terms of $\{\dot{r},\dot{\theta},\dot{\phi}\}$ is given by \cite{Wex1995,Konigsdorffer2005}
\begin{subequations}
\begin{align}
\label{EOM-fundamental-frame-r}
&\dot{r}^2=2(\varepsilon-\delta\varepsilon)+\frac{2m}{r}-\frac{h^2}{r^2},\\
\label{EOM-fundamental-frame-theta-varphi}
&\dot{\phi}=\frac{h}{r^2}\frac{\cos\Theta}{\sin^2\theta},\quad
\dot{\theta}^2=\left(\frac{h}{r^2}\sin\Theta\right)^2.
\end{align}
\end{subequations}
The angle $\Theta$ is obtained by solving the precession equation (\ref{precession-equation}). An equivalent but simpler method to determine $\theta$ and $\phi$-motion comes from the coordinate transformation between the fundamental and orbital frames. On the one hand, the relative position is expressed as $\hat{\bm{n}}=(\cos\psi,\sin\psi,0)$ in the orbital frame, and $\hat{\bm{n}}=(\sin\theta\cos\phi,\sin\theta\sin\phi,\cos\theta)$ in the fundamental frame. Applying the Euler rotation (\ref{transformation-from-fundamental-to-orbital-frame}), the equal expression of $\hat{\bm{n}}$ in the fundamental frame is
\begin{equation}
\label{n-fundamental-orbital-frame}
\hat{\bm{n}}
=\left(\begin{array}{c}
\cos\Upsilon\cos(\psi+\alpha)-\cos\Theta\sin\Upsilon\sin(\psi+\alpha)\\
\cos\Theta\cos\Upsilon\sin(\psi+\alpha)+\sin\Upsilon\cos(\psi+\alpha)\\
\sin\Theta\sin(\psi+\alpha)
\end{array}\right),
\end{equation}
Comparing the above two results of $\hat{\bm{n}}$, one finds \cite{Wex1995,Konigsdorffer2005}
\begin{equation}
\label{orbital-fundamental-frame-relationship-2}
\left\{\begin{array}{c}
\sin\theta\cos(\phi-\Upsilon)=\cos(\psi+\alpha)\\
\sin\theta\sin(\phi-\Upsilon)=\cos\Theta\sin(\psi+\alpha)\\
\cos\theta=\sin\Theta\sin(\psi+\alpha)
\end{array}\right.,
\end{equation}
The angles $\Theta$ and $\Upsilon$ are solved from the precession equation (\ref{precession-equation}), $\alpha$ is given by Eq.\,(\ref{alpha-equation}), and $\psi$ represents the motion on the orbital plane, which is given by a much simpler equation
\begin{equation}
\label{EOM-orbital-frame-psi}
\dot{\psi}=\frac{h}{r^2},
\end{equation}
depending only on the OAM magnitude $h$.

It should be noted that $\delta\varepsilon$ and $h$ in Eqs.\,(\ref{EOM-fundamental-frame-r},\,\ref{EOM-orbital-frame-psi}) depend on the spin vector, and they are not constants in the spin-precessing systems. To show this point, we express the SAM vector as
\begin{equation}
\label{spin-vector-in-orbital-frame}
\bm{S}_{A}=m_{A}^2\chi_{A}\left(
\sin\kappa_A\cos\psi_A,
\sin\kappa_A\sin\psi_A,
\cos\kappa_A\right)
\end{equation}
in the orbital frame, with $\chi_A$ being the dimensionless spin of the black holes. In terms of $\kappa_A$ and $\psi_A$, the dCS correction of energy is $\delta\varepsilon=\delta\varepsilon_{\rm SS}+\delta\varepsilon_{\rm MQ}$, where
\begin{subequations}
\begin{align}
&\begin{aligned}
\delta\varepsilon_{\rm SS}&=-\frac{25}{256}
\zeta\left(\frac{m}{r}\right)^3
\frac{\chi_1\chi_2}{\nu}\left\{\cos\kappa_1\cos\kappa_2
-\frac{1}{2}\sin\kappa_1\sin\kappa_2
\left[3\cos2(\psi-\bar{\psi})
+\cos\Delta\psi\right]\right\},
\end{aligned}\\
&\delta\varepsilon_{\rm MQ}=\frac{201}{3584}\zeta\left(\frac{m}{r}\right)^3
\sum_{A}
\frac{m^2}{m_A^2}\chi_A^2\left\{\cos^2\kappa_A
-\frac{1}{2}\sin^2\kappa_A \left[1+3\cos2(\psi-\psi_A)\right]\right\}.
\end{align}
\end{subequations}
The newly defined quantities are $\bar{\psi}\equiv(\psi_1+\psi_2)/2$ and $\Delta\psi\equiv\psi_1-\psi_2$. From the precession equation and TAM conservation, $\dd\bm{J}/\dd\bm{t}$, we can also derive the time derivative of $h$,  
\begin{equation}
\label{dotL-in-orbital-frame-psi}
\begin{aligned}
\dot{h}=\zeta\left(\frac{m}{r}\right)^3\left\{\frac{75}{256}\frac{\chi_1\chi_2}{\nu}\sin\kappa_1\sin\kappa_2\sin2(\psi-\bar{\psi})
-\frac{603}{1792}\sum_{A}\frac{m^2}{m_A^2}\chi_A^2\sin^2\kappa_A\sin2(\psi-\psi_A)\right\},
\end{aligned}
\end{equation}
implying that, in the spin-precessing case, $h$ is not a conserved quantity at 2PN order.

Two issues have arisen here. Firstly, Eqs.\,(\ref{EOM-fundamental-frame-r},\,\ref{EOM-orbital-frame-psi}) depend on the evolution of SAM vector, described by $\kappa_A$ and $\psi_A$, which is solved from the precession equation (\ref{precession-equation}). Meanwhile, the precession equation (\ref{precession-equation}) also depends on the relative position $\hat{\bm{n}}$. In other words, the precession and the acceleration equations are coupled. Subsection \ref{subsec:MSA} will introduce the multiple-scale analysis (MSA) to decouple them at different time scales. Secondly, Eq.\,(\ref{dotL-in-orbital-frame-psi}) shows that there is only one conserved quantity $\varepsilon$ in Eqs.\,(\ref{EOM-fundamental-frame-r},\,\ref{EOM-orbital-frame-psi}), which prevents us from defining the orbital semi-major axis and eccentricity as in the spin-aligned case. In Subsection \ref{subsec:AOAM}, we will introduce the period-averaged OAM magnitude as a new conserved quantity and further re-express Eq.\,(\ref{EOM-orbital-frame-psi}).

\subsection{\label{subsec:MSA}Multiple-scale analysis}

As discussed in the previous subsection, the acceleration and precession equations are coupled. However, the characteristic timescales are significantly distinguishable for the orbital motion and precession. One can estimate that $\tau_{\mathrm{orbit}}\sim|\bm{v}|/|\dot{\bm{v}}|\sim mv^{-3}$ from Eq.\,(\ref{acceleration-equation}), and $\tau_{\mathrm{precession}}\sim|\bm{S}_1|/|\dot{\bm{S}}_1|\sim mv^{-5}$ from Eq.\,(\ref{precession-equation}). $\tau_{\mathrm{precession}}$ is much larger than $\tau_{\mathrm{orbit}}$ within the PN framework. This inspires us to separate orbital motion and spin precession at different timescales. When investigating orbital motion, the SAM only undergoes slight changes, and the geometric quantities, such as $\kappa_A, \psi_A, \Theta, \Upsilon, \alpha$, related to the SAM vectors can be considered as constants within a few orbital periods. On the contrary, binary objects complete hundreds or even thousands of orbital motions within one precession period. The "details" of orbital motion in each period have no significant impact on the SAM evolution. Therefore, such impacts can be sufficiently described by the averaged orbital motion. The above discussion is the main strategy of the MSA. Chatziioannou \textit{et al} decoupled the precession equation through MSA and presented its analytical solution in terms of Jacobian elliptic function \cite{Chatziioannou2017}, while Loutrel \textit{et al} extended this analysis to dCS gravity within the circular-orbit approximation \cite{Loutrel2022}. This work mainly focuses on solving the decoupled acceleration equation in dCS gravity. The spin-related parameters, such as $\kappa_A$ and $\psi_A$, slowly evolve on longer timescales, bringing complicated low-frequency modulation to the gravitational waveform. In this work, we will treat them as constants temporarily, within the framework of MSA.

\subsection{\label{subsec:AOAM}Period-averaged orbital angular momentum magnitude}

As shown before, spin precession violates the OAM magnitude conservation at 2PN order. To provide sufficient conservation quantities for the parameterization of orbital motion, this subsection defines the period-averaged OAM magnitude. The solution to radial motion is approximated as
\begin{equation}
\label{r-psi-Newtonian-limit}
r=mj^2(1+a\cos V)^{-1},
\end{equation}
at the Newtonian order, where $j=h/m$ is the dimensionless OAM magnitude, and $a$ is the dimensionless magnitude of the Runge-Lenz vector at the Newtonian order, defined as
\begin{equation}
a=\sqrt{1+2j^2\varepsilon_{\rm N}}.
\end{equation}
The true anomaly is given by $V\equiv\psi-\psi_0$. Inserting Eq.\,(\ref{r-psi-Newtonian-limit}) into Eq.\,(\ref{dotL-in-orbital-frame-psi}), we get
\begin{equation}
\label{dotL-in-orbital-frame-psi-j}
\begin{aligned}
m\dot{j}=j^{-6}
\left(1+a\cos V\right)^{3}
&\times\zeta\Bigg\{\frac{75}{256}\frac{\chi_1\chi_2}{\nu}\sin\kappa_1\sin\kappa_2\sin2(V+\psi_0-\bar{\psi})\\
&\qquad\quad-\frac{603}{1792}\sum_{A}\frac{m^2}{m_A^2}\chi_A^2\sin^2\kappa_A\sin2(V+\psi_0-\psi_A)\Bigg\}.
\end{aligned}
\end{equation}
One can easily verify that 
\begin{equation}
\label{dotL-in-orbital-frame-average}
\frac{1}{T}\int_{0}^{T}\dot{j}\dd t=0,
\end{equation}
where $T$ is the orbital period. Eq.\,(\ref{dotL-in-orbital-frame-average}) inspires us that although the OAM magnitude $j$ varies over time, it is only a periodic oscillation. The oscillation amplitude is a 2PN perturbation relative to the averaged value. We formally define the period-averaged OAM magnitude $\bar{j}$ as \cite{Gergely1999,Gergely2000,Gergely2003}
\begin{equation}
\label{L-average-in-orbital-frame-def}
\bar{j}\equiv\frac{1}{2\pi}\int_{\psi}^{\psi+2\pi}j(\psi')\dd\psi'.
\end{equation}
It is verified that $\bar{j}$ is conserved, and $j(t)=\bar{j}+\delta j(t)$, with $\delta j(t)$ being tiny periodic oscillation around $0$. Let us define the initial value of OAM magnitude as $j_0\equiv j(\psi(t)=0)$, and integrate $\dot{j}$ to give
\begin{equation}
\label{L-in-orbital-frame-L0-deltaL}
j(V)=j_0+\int_0^{\psi}\dot{j}\frac{\dd t}{\dd\psi'}\dd\psi'=j_0+\delta\tilde{j}(V).
\end{equation}
The correction $\delta\tilde{j}(V)$ is at 2PN and $\mathcal{O}(\zeta)$ order, showing $j(V=0)=j(V=2\pi)=0$. Inserting Eq.\,(\ref{L-in-orbital-frame-L0-deltaL}) into the definition of $\bar{j}$ (\ref{L-average-in-orbital-frame-def}) and performing the integration, we solve the relationship between $\bar{j}$ and $j_0$. Replacing the $j_0$ in Eq.\,(\ref{L-in-orbital-frame-L0-deltaL}) by $\bar{j}$, we obtain
\begin{equation}
j(V)=\bar{j}+\delta j,\quad
\delta j=\delta j_{\rm SS}+\delta j_{\rm MQ},
\end{equation}
where the leading-order dCS corrections are 
\begin{subequations}
\begin{align}
&\begin{aligned}
\delta j_{\rm SS}&=-\frac{25}{512}
\frac{\zeta}{\bar{j}^3}
\frac{\chi_1\chi_2}{\nu}\sin\kappa_1\sin\kappa_2
\Big\{3\cos2(V+\psi_0-\bar{\psi})\\
&\quad+\bar{a}\Big[3\cos(V+2(\psi_0-\bar{\psi}))+
\cos(3V+2(\psi_0-\bar{\psi}))\Big]\Big\},
\end{aligned}\\
&\begin{aligned}
\delta j_{\rm MQ}&=\frac{201}{7168}
\frac{\zeta}{\bar{j}^3}
\sum_A\frac{m^2}{m_{A}^2}\chi_A^2\sin^2\kappa_A
\Big\{3\cos2(V+\psi_0-\psi_A)\\
&\quad+\bar{a}\Big[3\cos(V+2(\psi_0-\psi_A))+
\cos(3V+2(\psi_0-\psi_A))\Big]\Big\},
\end{aligned}
\end{align}
\end{subequations}
with $\bar{a}\equiv(1+2\bar{j}^2\varepsilon_{\rm N})^{1/2}$. In terms of $\bar{j}$ and $\delta j$, we re-express Eqs.\,(\ref{EOM-fundamental-frame-r},\,\ref{EOM-orbital-frame-psi}) as
\begin{equation}
\label{dot-r}
\dot{r}^2=2(\varepsilon-\delta\varepsilon)+\frac{2m}{r}-\left(\frac{m}{r}\right)^2\left(\bar{j}^2+2\bar{j}\delta j\right),
\end{equation}
and
\begin{equation}
\label{dot-psi}
m\dot{\psi}=\bar{j}\left(\frac{m}{r}\right)^2
\left(1+\frac{\delta j}{\bar{j}}\right).
\end{equation}
Within the MSA framework, only three time-varying quantities, $\{r,\psi,V\}$, are involved in Eqs.\,(\ref{dot-r},\,\ref{dot-psi}), and other spin-related geometric quantities, like $\kappa_A, \psi_A$ are seen as constants. This completes all the preparations for the QK parameterization. In GR, the acceleration equation of an eccentric and precessing binary can be written in a similar form of Eqs.\,(\ref{dot-r},\,\ref{dot-psi}), but with different correction coefficients $\delta j$ and $\delta\varepsilon$.

\section{\label{sec:QK}Quasi-Keplerian parameterization}
The original parameterization scheme for Eqs.\,(\ref{dot-r},\,\ref{dot-psi}) is provided by Gergely \textit{et al} \cite{Gergely1999,Gergely2003}. However, this parameterization is singular when $\bar{a}$ tends towards $0$, causing the QK solution to fail at the zero-eccentricity limit. Klein \textit{et al} improved the above scheme in GR to avoid such a singularity \cite{Klein2010,Klein2018}. We extend Klein's scheme to dCS gravity, expressing the parameterization formally as
\begin{equation}
\label{r-sol-0}
\begin{aligned}
r&=\frac{m}{\xi}\Bigg\{(1-e_r\cos u)+f^{\rm(SS)}_{r}\cos(2V+2\psi_0-\psi_1-\psi_2)\\
&\qquad\qquad\qquad\qquad+\sum_{A}f^{\rm(MQ)}_{r,A}\cos[2V+2(\psi_0-\psi_A)]\Bigg\},
\end{aligned}
\end{equation}
\begin{equation}
\label{psi-sol-0}
\begin{aligned}
\frac{2\pi}{K}\psi&=2\tan^{-1}\left\{\sqrt{\frac{1+e_\psi}{1-e_\psi}}\tan\left(\frac{u}{2}\right)\right\}+\sum_{n=1}^{2}\Bigg\{f^{\rm(SS)}_{\psi,n}\sin(nV+2\psi_0-\psi_1-\psi_2)\\
&\qquad\qquad\qquad\qquad\qquad\qquad+\sum_{A}f^{\rm(MQ)}_{\psi,n,A}\cos[nV+2(\psi_0-\psi_A)]\Bigg\},
\end{aligned}
\end{equation}
\begin{equation}
\label{t-sol-0}
\frac{2\pi}{T}t(u)=u-e_t\sin u.
\end{equation}
In Eqs.\,(\ref{r-sol-0},\,\ref{psi-sol-0},\,\ref{t-sol-0}), the binary separation, azimuth angle, and coordinate time are written as a set of functions about true anomaly $V$, and eccentric anomaly $u$, relating $V$ by
\begin{equation}
\label{eccentric-anomaly}
u\equiv2\arctan\left[\sqrt{\frac{1-e_r}{1+e_r}}
\tan\left(\frac{V}{2}\right)\right].
\end{equation}
Additionally, a series of parameters is introduced. $\xi\equiv m/a_r$ with $a_r$ being the semi-major axis of the binary orbit. $e_r, e_\psi, e_t$ denote the ``radial", ``azimuth", and ``time" eccentricity. $T$ and $K$ are ``time" and ``azimuth” period. $f^{\rm(SS)}_{r},f^{\rm(MQ)}_{r,A},f^{\rm(SS)}_{\psi,n},f^{\rm(MQ)}_{\psi,n,A}$ are correction coefficients relating with the spin precession. Such a parameterization scheme involves a total of 15 parameters, but only two of them are independent. This is because the orbital motion is completely determined by the conserved quantities $\varepsilon$ and $\bar{j}$. Following our previous works \cite{ZhaoLi2024}, we choose $\xi$ and $e_r$ as free parameters, and $\varepsilon, \bar{j}$ are expressed as
\begin{subequations}
\begin{align}
\varepsilon&=-\frac{\xi}{2}\left\{1-\left(\frac{\xi^2}{1-e_r^2}\right)\delta\varpi\right\},\\
\bar{j}&=\sqrt{\frac{1-e_r^2}{\xi}}\left\{1+\frac{1}{2}\left(\frac{\xi}{1-e_r^2}\right)^2(3+e_r^2)\delta\varpi\right\},
\end{align}
\end{subequations}
providing the relation between conservation and orbital elements. The dCS correction $\delta\varpi$ is 
\begin{equation}
\begin{aligned}
\delta\varpi&\equiv\zeta\Bigg\{\frac{25}{512}\frac{\chi_1\chi_2}{\nu}\left(2\cos\kappa_1\cos\kappa_2-\sin\kappa_1\sin\kappa_2\cos\Delta\psi\right)\\
&\qquad\qquad\qquad-\frac{201}{14336}\sum_{A}\frac{m^2}{m_A^2}\chi_A^2(1+3\cos2\kappa_A)\Bigg\},
\end{aligned}
\end{equation}
which returns to the definition given by Ref.\,\cite{ZhaoLi2024} in spin-aligned limit, where $\kappa_A=0$. The other two eccentricities are
\begin{equation}
\label{et-ephi-er}
e_{t}=e_{r}\left(1
-\frac{1}{3}\delta\varpi
\frac{\xi^2}{1-e_r^2}\right),\quad
e_{\phi}=e_{r}\left(1
+\frac{1}{3}\delta\varpi
\frac{\xi^2}{1-e_r^2}\right),
\end{equation}
and two different orbital periods, ``time“ and ``azimuth” period, are
\begin{equation}
\label{T-K}
T=2\pi\frac{m}{\xi^{3/2}}\left(
1+\frac{1}{2}\delta\varpi
\frac{\xi^2}{1-e_r^2}\right),\quad
K=2\pi\left[1+\delta\varpi\left(\frac{\xi}{1-e_r^2}\right)^2\right],
\end{equation}
in terms of fundamental elements $\xi$ and $e_r$. The ``time” period is defined as the time it takes for the objects to pass through the pericenter twice, and the ``azimuth" period is defined as the time it takes for the objects to pass through the starting point of the right ascension twice. The precession-related parameters are
\begin{subequations}
\begin{align}
f_{r}^{\rm(SS)}&=-\frac{25}{512}\zeta\left(\frac{\xi^2}{1-e_r^2}\right)\frac{\chi_1\chi_2}{\nu}\sin\kappa_1\sin\kappa_2,\\
f_{r,A}^{\rm(MQ)}&=\frac{201}{7168}\zeta\left(\frac{\xi^2}{1-e_r^2}\right)\frac{m^2}{m_A^2}\chi_A^2\sin^2\kappa_A,\\
f_{\psi,1}^{\rm(SS)}&=-\frac{25}{256}\zeta e_r\left(\frac{\xi}{1-e_r^2}\right)^2\frac{\chi_1\chi_2}{\nu}\sin\kappa_1\sin\kappa_2,\\
f_{\psi,1,A}^{\rm(MQ)}&=\frac{201}{3584}\zeta e_r\left(\frac{\xi}{1-e_r^2}\right)^2\frac{m^2}{m_2^2}\chi_A^2\sin^2\kappa_A,\\
f_{\psi,2}^{\rm(SS)}&=-\frac{25}{1024}\zeta\left(\frac{\xi}{1-e_r^2}\right)^2\frac{\chi_1\chi_2}{\nu}\sin\kappa_1\sin\kappa_2,\\
f_{\psi,2,A}^{\rm(MQ)}&=\frac{201}{14336}\zeta\left(\frac{\xi}{1-e_r^2}\right)^2\frac{m^2}{m_A^2}\chi_A^2\sin^2\kappa_A.
\end{align}
\end{subequations}
It should be pointed out that $f_{r}^{\rm(SS)}$ and $f_{r,A}^{\rm(MQ)}$ are non-vanishing when eccentricity $e_r$ approaches zero, implying that $r$ is not a constant even for zero-eccentricity case. A periodic oscillation proportional to $\cos2V$ is attached to the $r$-solution. This is caused by spin precession, and then the eccentricity can only be regarded as a physical parameter related to conserved quantities and cannot be used to understand the orbital geometry. Therefore, this work will distinguish between zero-eccentricity and conventional circular orbits.

As a summary, we collect the expression of $r$ and $\psi$ below and replace all $u$ with the true anomaly $V$,
\begin{equation}
\label{radial-parameterization}
\begin{aligned}
\frac{m}{r}&=\frac{\xi}{1-e_r^2}(1+e_r\cos V)
+\frac{\zeta}{2}\left(\frac{\xi}{1-e_r^2}\right)^3(1+e_r\cos V)^2\\
&\quad\times\left\{\frac{25}{256}\frac{\chi_1\chi_2}{\nu}\sin\kappa_1\sin\kappa_2
\cos\left[2(1+\beta)V-2\bar{\psi}\right]\right.\\
&\quad\left.-\frac{201}{3584}\sum_{A}\frac{m^2}{m_A^2}\chi_A^2\sin^2\kappa_A
\cos\left[2(1+\beta)V-2\psi_A\right]\right\},
\end{aligned}
\end{equation}
\begin{equation}
\label{psi-parameterization}
\begin{aligned}
\psi&=(1+\beta)V
+\delta\varpi\left(\frac{\xi}{1-e_r^2}\right)^2e_r\sin V+\frac{\zeta}{4}\left(\frac{\xi}{1-e_r^2}\right)^2\\
&\times\left\{-\frac{25}{256}\frac{\chi_1\chi_2}{\nu}\sin\kappa_1\sin\kappa_2
\Big\{4e_r\sin\left[(1+2\beta)V-2\bar{\psi}\right]
+\sin\left[2(1+\beta)V-2\bar{\psi}\right]\Big\}\right.\\
&\left.+\sum_{A}\frac{201}{3584}\frac{m^2}{m_A^2}\chi_A^2\sin^2\kappa_A
\Big\{4e_r\sin\left[(1+2\beta)V-2\psi_A\right]
+\sin\left[2(1+\beta)V-2\psi_A\right]\Big\}\right\}.
\end{aligned}
\end{equation}
$\psi_0$ is approximated as $\psi_0=\beta V$ at the leading order, with
\begin{equation}
\beta\equiv\delta\varpi\left(\frac{\xi}{1-e_r^2}\right)^2
\end{equation}
being the pericenter advance rate. 

When discussing spin-aligned binaries in our previous work, we mentioned the doubly periodic structure of eccentric motion in the PN framework, where the binary objects pass through an azimuth angle of more than $2\pi$ within a ``time" period \cite{ZhaoLi2024}. This is the well-known pericenter advance effect, entering the 2PN order in our investigation. The spin precession leads to a more complicated periodic structure by introducing richer long-period modulation. For instance, the nutation and azimuthal drift of SAM relative, described by $\kappa_A$ and $\psi_A$. However, these effects are beyond the scope of this work, and we have regarded them as constants within the MSA framework. More comprehensive discussions and calculations will be performed in our future work.

\section{\label{sec:GW}Gravitational waveform: conservative sector}
After dynamics modeling, the scalar and gravitational radiation are routinely calculated by inserting the relative position $\hat{\bm{n}}$, velocity vector $\bm{v}$, and SAM vectors $\bm{S}_A$ into Eqs.\,(\ref{scalar-radiation},\,\ref{tensor-radiation}). The scalar waveform is given from Eq.\,(\ref{scalar-radiation}) by 
\begin{equation}
\label{scalar-waveform}
\begin{aligned}
\vartheta&=\frac{5}{128}\frac{\bar{\alpha}}{\bar{\beta}m^2}\left(\frac{\xi}{1-e_r^2}\right)^2\frac{m_2}{m}\frac{\chi_1}{\nu}\Big\{\vartheta_{\rm(I)}\sin^2\iota\sin\kappa_1\\
&\qquad+\vartheta_{\rm(II)}(3-\cos^2\iota)\sin\kappa_1
+\vartheta_{\rm(III)}\sin2\iota\cos\kappa_1\Big\}-(1\leftrightarrow2),
\end{aligned}
\end{equation}
where
\begin{subequations}
\begin{align}
&\begin{aligned}
\vartheta_{\rm(I)}&=-4e_r\cos(\beta V+2 \tilde{\alpha}+\psi_1)
-2(2+e_r^2)\cos[(1+\beta)V+2\tilde{\alpha}+\psi_1]\\
&-e_r^2\cos[(1-\beta)V-2\tilde{\alpha}-\psi_1]
-4e_r\cos{[(2+\beta)V+2\tilde{\alpha}+\psi_1]}\\
&-e_r^2\cos[(3+\beta)V+2\tilde{\alpha}+\psi_1],
\end{aligned}\\
&\begin{aligned}
\vartheta_{\rm(II)}&=4e_r\cos(\beta V-\psi_1)
+2(2+e_r^2)\cos{[(1+\beta)V-\psi_1]}
+e_r^2\cos{[(1-\beta)V+\psi_1]}\\
&+4e_r\cos{[(2+\beta)V-\psi_1]}
+e_r^2\cos{[(3+\beta)V-\psi_1]},
\end{aligned}\\
&\begin{aligned}
\vartheta_{\rm(III)}&=4e_r\sin(\beta V+\tilde{\alpha})
+2(2+e_r^2)\sin[(1+\beta)V+\tilde{\alpha}]
-e_r^2\sin[(1-\beta)V-\tilde{\alpha}]\\
&+4e_r\sin[(2+\beta)V+\tilde{\alpha}]
+e_r^2\sin[(3+\beta)V+\tilde{\alpha}],
\end{aligned}
\end{align}
\end{subequations}
and $\tilde{\alpha}\equiv\alpha+\omega+\Upsilon$. The $(1\leftrightarrow2)$ at the end of Eq.\,(\ref{scalar-waveform}) mean exchanging all of the subscript $1$ and $2$, e.g., replacing $\kappa_1$ with $\kappa_2$, $\chi_1$ with $\chi_2$,
and $m_2$ with $m_1$.

The gravitational waveform is evaluated from Eq.\,(\ref{tensor-radiation}), and the GW polarizations are obtained by performing a rotation and TT projection from $\xi_{ij}$. We denote the GW metric in the TT gauge as $\xi_{ij}^{\rm TT}$, and the plus and cross modes are $\xi_{+}=\xi_{11}^{\rm TT},\xi_{\times}=\xi_{12}^{\rm TT}$. In the spin-aligned case, $\xi_{+,\times}$ can be fully separated into the Newtonian waveform and the dCS correction as
\begin{equation}
\label{xi-def}
\xi_{+,\times}=\xi_{+,\times}^{(0)}+\delta\xi_{+,\times},
\end{equation}
$\xi_{+,\times}^{(0)}$ is completely same as that in GR. However, in the spin-precessing case, the nutation and azimuthal drift of the OAM vector bring additional amplitude and phase modulation, carrying the parity-violating modification. 

We insist on expressing the full gravitational waveform in Eq.\,(\ref{xi-def}). However, it should be noted that $\xi_{+,\times}^{(0)}$ is no longer entirely the GR waveforms, although they still appear to be of $\mathcal{O}(\zeta^0)$ order. $\xi_{+,\times}^{(0)}$ contains the amplitude and phase modulations, encoding the dCS modification. The phase modulation is represented by angles $\Upsilon$ and $\alpha$, and the amplitude modulation is described by nutation angle $\Theta$, which is of order $|\bm{S}_A|/|\bm{L}|\sim\mathcal{O}(v)$. Thus we can expand $\xi_{+,\times}^{(0)}$ in terms of small parameter $\sin\Theta$ as
\begin{equation}
\label{xi-0}
\xi^{(0)}_{+,\times}=\Sigma^{(0)}_{+,\times}
+\sin\Theta\Sigma^{(1/2)}_{+,\times}
+\sin^2\Theta\Sigma^{(1)}_{+,\times}
+\sin^3\Theta\Sigma^{(3/2)}_{+,\times}
+\sin^4\Theta\Sigma^{(2)}_{+,\times},
\end{equation}
where the waveforms at each PN order are given in Appendix \ref{Sigma-expression}. The dCS correction, which is proportional to $\zeta$ is given by
\begin{equation}
\delta\xi_{+,\times}=\delta\xi_{+,\times}^{\rm(SS)}+\delta\xi_{+,\times}^{\rm(MQ)},
\end{equation}
where the SS-coupling and MQ-coupling terms are
\begin{subequations}
\label{delta-xi-SS}
\begin{align}
\delta\xi^{\rm(SS)}_{+}&=\frac{75}{512}\zeta\left(\frac{\xi}{1-e_r^2}\right)^3\frac{\chi_1\chi_2}{\nu}\sum_{k=0}^7\left[\mathcal{A}^{+}_{k}\cos(kV)+\mathcal{B}^{+}_{k}\sin(kV)\right],\\
\delta\xi^{\rm(SS)}_{\times}&=\frac{75}{512}\zeta\left(\frac{\xi}{1-e_r^2}\right)^3\frac{\chi_1\chi_2}{\nu}\sum_{k=0}^7\left[\mathcal{A}^{\times}_{k}\cos(kV)+\mathcal{B}^{\times}_{k}\sin(kV)\right],
\end{align}
\end{subequations}
and
\begin{subequations}
\label{delta-xi-MQ}
\begin{align}
\delta\xi^{\rm(MQ)}_{+}&=\frac{201}{1792}\zeta\left(\frac{\xi}{1-e_r^2}\right)^3\sum_{A}\frac{m^2}{m_A^2}\chi_A^2
\sum_{k=0}^7\left[\mathcal{C}^{+}_{A,k}\cos(kV)+\mathcal{D}^{+}_{A,k}\sin(kV)\right],\\
\delta\xi^{\rm(MQ)}_{\times}&=\frac{201}{1792}\zeta\left(\frac{\xi}{1-e_r^2}\right)^3\sum_{A}\frac{m^2}{m_A^2}\chi_A^2\sum_{k=0}^7\left[\mathcal{C}^{\times}_{A,k}\cos(kV)+\mathcal{D}^{\times}_{A,k}\sin(kV)\right].
\end{align}
\end{subequations}
The coefficients $\mathcal{A}_k^{+,\times}, \mathcal{B}_k^{+,\times}, \mathcal{C}_{A,k}^{+,\times}, \mathcal{D}_{A,k}^{+,\times}$ are listed in Appendix \ref{app-SS} and \ref{app-MQ}, respectively. It should be emphasized again that these coefficients are not constants, but evolve over longer time scales and ultimately bring low-frequency phase modulation into the gravitational waveform. This work temporarily considers them as constants within the MSA framework. The variables involved in Eq.\,(\ref{xi-def}), such as nutation angle $\Theta$, azimuthal drift $\Upsilon, \alpha$, and the direction of SAM vectors, $\kappa_A, \psi_A$, can be obtained once the evolution of the $\bm{S}_A$ vectors is solved \cite{Chatziioannou2017,Loutrel2022}.

\section{\label{sec:RR}Gravitational radiation and orbital decay}

\subsection{Energy and angular momentum loss}

Gravitational radiation carries energy and angular momentum from the wave sources. As the binding energy and angular momentum of BBH dissipate, the semi-major axis and eccentricity gradually decrease, and the orbital frequency increases until the two black holes collide and merge, ultimately forming an isolated black hole. The correction induced by dCS gravity to the gravitational waveform and scalar radiation will accelerate/decelerate orbital decay and circularization. The energy loss of the BBH system is separated into the tensor and scalar sectors.
\begin{equation}
\label{enery-flux-def}
\mathcal{F}=\mathcal{F}_{T}+\mathcal{F}_{S},
\end{equation}
where the contribution of GW is 
\begin{equation}
\label{enery-flux-tensor-def}
\mathcal{F}_{T}=\frac{1}{32\pi}R^2\left[\oint_{\partial\Omega}
\langle\dot{h}^{\rm TT}_{jk}\dot{h}^{\rm TT}_{jk}\rangle\dd\Omega\right],
\end{equation}
and that of the scalar field is
\begin{equation}
\label{enery-flux-scalar-def}
\mathcal{F}_{S}=\bar{\beta} R^2\oint_{\partial\Omega}\langle\dot{\vartheta}^2\rangle\dd\Omega.
\end{equation}
Inserting Eqs.\,(\ref{scalar-waveform},\,\ref{xi-def}) into Eq.\,(\ref{enery-flux-def}), we obtain
\begin{equation}
\label{energy-flux-res}
\begin{aligned}
\mathcal{F}&=\frac{32}{5}\frac{\nu^2\xi^5}{(1-e_r^2)^{7/2}}\Bigg\{
\left(1+\frac{73}{24}e_r^2+\frac{37}{96}e_r^4\right)\\
&+\zeta\left(\frac{\xi}{1-e_r^2}\right)^2\Bigg\{\frac{\chi_1\chi_2}{\nu}\Bigg[
\left(\frac{10775}{12288}
+\frac{137825}{24576}e_r^2
+\frac{133925}{32768}e_r^4
+\frac{49975}{196608}e_r^6
\right)
\cos\kappa_1\cos\kappa_2\\
&-\left(\frac{4325}{24576}
+\frac{47675}{49152}e_r^2
+\frac{50375}{65536}e_r^4
+\frac{24325}{393216}e_r^6
\right)\sin\kappa_1\sin\kappa_2\cos\Delta\psi\\
&-e_r^2\left(\frac{118825}{32768}
+\frac{351775}{98304}e_r^2
+\frac{112625}{524288}e_r^4
\right)
\sin\kappa_1\sin\kappa_2\cos2(\bar{\psi}+\beta V)\Bigg]\\
&-\sum_{A}\frac{m^2}{m_A^2}\chi_A^2\Bigg[
\left(\frac{23869}{344064}
+\frac{470395}{688128}e_r^2
+\frac{422759}{917504}e_r^4
+\frac{23903}{1835008}e_r^6\right)\\
&+\left(\frac{49815}{114688}
+\frac{582273}{229376}e_r^2
+\frac{1731951}{917504}e_r^4
+\frac{244119}{1835008}e_r^6
\right)\cos\kappa_A^2\\
&-e_r^2\left(\frac{567591}{458752}
+\frac{563707}{458752}e_r^2
+\frac{535167}{7340032}e_r^4
\right)\sin\kappa_A^2\cos2(\psi_A+\beta V)\Bigg]\Bigg\}\Bigg\}.
\end{aligned}
\end{equation}
up to $\mathcal{O}(\zeta)$ order.
We consider the slowly-evolving variables $\psi_0=\beta V, \kappa_A, \psi_A$ as constants. There are only Newtonian terms and 2PN dCS correction involved in Eq.\,(\ref{energy-flux-res}), and there is no contribution from correction terms in Eq.\,(\ref{xi-0}). This is because the spin precession affects the gravitational waveforms through Euler rotation, and $\Theta, \Upsilon$ degenerates with the angular distribution of radiated energy flux, as described by $\iota, \omega$.

From the definition of angular momentum loss,
\begin{equation}
\label{angular-momentum-flux-def}
\bm{\mathcal{J}}=\bm{\mathcal{J}}_{T}+\bm{\mathcal{J}}_{S},
\end{equation}
where the contributions from GWs and scalar radiation are
\begin{equation}
\label{angular-momentum-flux-tensor-def}
\mathcal{J}_T^k
=\frac{1}{32\pi}R^2
\int\langle\tau_{T}^k\rangle \dd\Omega,\quad
\tau_{T}^k=\epsilon^{ijk}
(2h_{il}^{\mathrm{TT}} \dot{h}_{jl}^{\mathrm{TT}}
-\dot{h}_{lm}^{\mathrm{TT}}
x_i\partial_jh_{lm}^{\mathrm{TT}}),
\end{equation}
and
\begin{equation}
\label{angular-momentum-flux-scalar-def}
\mathcal{J}^{k}_S
=-\bar{\beta}R^2\int
\langle\tau_{S}^k\rangle \dd\Omega,\quad
\tau_{S}^k=-\epsilon_{ijk}\dot{\vartheta}
x_i\partial_{j}\vartheta,
\end{equation}
respectively. We care about the radiation reaction on the orbital evolution; only the magnitude of OAM loss needs to be calculated. We denote the TAM loss as $\dd\bm{J}/\dd t$, consisting of the losses of OAM and SAM, i.e., $(\dd/\dd t)(\bm{L}+\bm{S}_1+\bm{S}_2)$. Ref.\,\cite{Gergely1999} has proven that the SAM contribution enters the orders higher than 2PN; we will not consider this sector in this work. The magnitude of OAM loss is $\dd L/\dd t=\hat{\bm{L}}\cdot(\dd\bm{L}/\dd t)$, and the magnitude of TAM loss is $\dd J/\dd t=\hat{\bm{J}}\cdot(\dd\bm{J}/\dd t)\propto\hat{\bm{J}}\cdot\hat{\bm{L}}(\dd L/\dd t)\propto\cos\Theta(\dd L/\dd t)$. Therefore, we have $\mathcal{L}\equiv\dd L/\dd t=(\cos\Theta)^{-1}(\dd J/\dd t)=(\cos\Theta)^{-1}\mathcal{J}$. After inserting Eqs.\,(\ref{scalar-waveform},\,\ref{xi-def}) into the definition (\ref{angular-momentum-flux-def}) and tedious calculations, we arrive at
\begin{equation}
\label{angular-momentum-flux-res}
\begin{aligned}
\mathcal{L}&=\frac{32}{5}\frac{m\nu^2\xi^{7/2}}{(1-e_r^2)^{2}}\Bigg\{
\left(1+\frac{7}{8}e_r^2\right)\\
&+\zeta\left(\frac{\xi}{1-e_r^2}\right)^2\Bigg\{\frac{\chi_1\chi_2}{\nu}\Bigg[
\left(\frac{8975}{12288}
+\frac{1125}{512}e_r^2
+\frac{16775}{32768}e_r^4\right)
\cos\kappa_1\cos\kappa_2\\
&-\left(\frac{475}{6144}
+\frac{1925}{8192}e_r^2
+\frac{2425}{16384}e_r^4
\right)\sin\kappa_1\sin\kappa_2\cos\Delta\psi\\
&-e_r^2\left(\frac{7125}{8192}
+\frac{825}{4096}e_r^2\right)
\sin\kappa_1\sin\kappa_2\cos2(\bar{\psi}+\beta V)\Bigg]\\
&-\sum_{A}\frac{m^2}{m_A^2}\chi_A^2\Bigg[\left(\frac{10679}{86016}
+\frac{42515}{114688}e_r^2
+\frac{4999}{229376}e_r^4\right)\\
&+\left(\frac{7261}{24576}
+\frac{102257}{114688}e_r^2
+\frac{144895}{458752}e_r^4
\right)\cos\kappa_A^2\\
&-e_r^2\left(\frac{31959}{114688}
+\frac{9045}{114688}e_r^2
\right)\sin\kappa_A^2\cos2(\psi_A+\beta V)\Bigg]\Bigg\}\Bigg\},
\end{aligned}
\end{equation}
up to the leading-order dCS correction. Similar to the result of energy loss (\ref{energy-flux-res}), the OAM loss only includes Newtonian terms and 2PN correction, without contribution from spin-related higher-order corrections in Eq.\,(\ref{xi-0}).

\subsection{The evolution of elements}
After presenting the energy and angular momentum loss, the radiation-reaction evolution of orbital elements is given by the balance equation
\begin{equation}
\frac{\dd E}{\dd t}=-\mathcal{F},\quad
\frac{\dd\bar{L}}{\dd t}=-\mathcal{L},
\end{equation}
with $\bar{L}=\mu\bar{h}=\mu m\bar{j}$. Regarding the elements $\xi$ and $e_r$ as functions of time and solving the balance equation using Eq.\,(\ref{energy-flux-res},\,\ref{angular-momentum-flux-res}), one can obtain the time derivative of elements.
\begin{equation}
\label{dx-dt}
\begin{aligned}
m\frac{\dd x}{\dd t}&=\frac{64}{5}\frac{\nu x^5}{(1-e_r^2)^{7/2}}\Bigg\{
\left(1+\frac{73}{24}e_r^2+\frac{37}{96}e_r^4\right)\\
&+\zeta\left(\frac{x}{1-e_r^2}\right)^2\Bigg\{\frac{\chi_1\chi_2}{\nu}\Bigg[
\left(\frac{11975}{12288}
+\frac{102725}{24576}e_r^2
+\frac{211075}{98304}e_r^4
+\frac{4325}{65536}e_r^6\right)
\cos\kappa_1\cos\kappa_2\\
&+\left(-\frac{5525}{24576}
-\frac{12575}{49152}e_r^2
+\frac{39575}{196608}e_r^4
+\frac{4225}{131072}e_r^6\right)\sin\kappa_1\sin\kappa_2\cos\Delta\psi\\
&-e_r^2\left(\frac{118825}{32768}
+\frac{351775}{98304}e_r^2
+\frac{112625}{524288}e_r^4\right)
\sin\kappa_1\sin\kappa_2\cos2(\bar{\psi}+\psi_0)\Bigg]\\
&-\sum_{A}\frac{m^2}{m_A^2}\chi_A^2\Bigg[\left(\frac{14221}{344064}
+\frac{752599}{688128}e_r^2
+\frac{133405}{131072}e_r^4
+\frac{123063}{1835008}e_r^6\right)\\
&+\left(\frac{59463}{114688}
+\frac{42867}{32768}e_r^2
+\frac{28389}{131072}e_r^4
-\frac{7623}{262144}e_r^6
\right)\cos\kappa_A^2\\
&-e_r^2\left(\frac{567591}{458752}
+\frac{563707}{458752}e_r^2
+\frac{535167}{7340032}e_r^4
\right)\sin\kappa_A^2\cos2(\psi_A+\beta V)\Bigg]\Bigg\}\Bigg\},
\end{aligned}
\end{equation}
and
\begin{equation}
\label{de-dt}
\begin{aligned}
m\frac{\dd e_r}{\dd t}&=-\frac{304}{15}\frac{\nu x^4 e_r}{(1-e_r^2)^{5/2}}\Bigg\{
\left(1+\frac{121}{304}e_r^2\right)\\
&+\zeta\left(\frac{x}{1-e_r^2}\right)^2\Bigg\{\frac{\chi_1\chi_2}{\nu}\Bigg[
\left(\frac{83825}{77824}
+\frac{180275}{155648}e_r^2
+\frac{56825}{622592}e_r^4\right)
\cos\kappa_1\cos\kappa_2\\
&+\frac{1}{e_r^2}\left(-\frac{625}{77824}
-\frac{21975}{155648}e_r^2
+\frac{38675}{622592}e_r^4
+\frac{11275}{1245184}e_r^6
\right)\sin\kappa_1\sin\kappa_2\cos\Delta\psi\\
&-\left(\frac{270975}{311296}
+\frac{417475}{311296}e_r^2
+\frac{654675}{4980736}e_r^4\right)
\sin\kappa_1\sin\kappa_2\cos2(\bar{\psi}+\beta V)\Bigg]\\
&+\sum_{A}\frac{m^2}{m_A^2}\chi_A^2\Bigg[\frac{1}{e_r^2}\left(\frac{625}{155648}-
\frac{445055}{2179072}e_r^2
-\frac{4012101}{8716288}e_r^4
-\frac{705885}{17432576}e_r^6\right)\\
&-\frac{1}{e_r^2}\left(\frac{625}{155648}
+\frac{903631}{2179072}e_r^2
+\frac{1790223}{8716288}e_r^4
+\frac{208641}{17432576}e_r^6
\right)\cos\kappa_A^2\\
&+\left(\frac{69435}{229376}
+\frac{1966089}{4358144}e_r^2
+\frac{3342141}{69730304}e_r^4
\right)\sin\kappa_A^2\cos2(\psi_A+\beta V)\Bigg]\Bigg\}\Bigg\}.
\end{aligned}
\end{equation}
The variable $x$ used in Eqs.\,(\ref{dx-dt}, \ref{de-dt}) is the dimensionless orbital frequency, defined as
\begin{equation}
x\equiv(m\Omega)^{2/3},
\end{equation}
with $\Omega$ being orbital frequency,
\begin{equation}
\label{Omega}
\Omega\equiv\frac{K}{T}=\frac{\xi^{3/2}}{m}
\left[1+\frac{\delta\varpi}{2}\left(\frac{\xi}{1-e_r^2}\right)^2(1+e_r^2)\right].
\end{equation}
Eq.\,(\ref{dx-dt}) indicates that, in the PN approximation, the orbital frequency increases with GW radiation until the approximation condition $x\ll1$ is broken. As is known, the eccentricity eventually tends to zero in GR, and the zero-eccentricity orbit is eventually stable in the BBH system under radiation reaction. However, after introducing dCS modification, the evolution equation of eccentricity (\ref{de-dt}) exhibits an interesting phenomenon; there is a factor proportional to $1/e_r$ on the right-hand side, such that $\dd e_r/\dd t$ diverges when $e_r\rightarrow0$, which is physically unacceptable. This indicates that in the spin-precession case, zero-eccentricity orbits cannot exist stably under radiation reaction in dCS gravity.

This effect is caused by monopole scalar radiation. Whether in GR and dCS gravity, GW is generated from mass quadrupole momentum and corresponding higher-order corrections, carrying both energy and angular momentum, and the identity $\mathcal{F}_T=\Omega\mathcal{L}_T$ is valid as the orbital eccentricity approaches zero. This ensures that the eccentric orbit evolves into a quasi-circular orbit under radiation reaction and forms a stable state. In dCS gravity, the scalar radiation (\ref{scalar-radiation}) contains two terms: monopole and quadrupole radiations. We split the energy and angular momentum losses into $\mathcal{F}_S=\mathcal{F}_{\rm mon}+\mathcal{F}_{\rm quad},\,\mathcal{L}_S=\mathcal{L}_{\rm mon}+\mathcal{L}_{\rm quad}$, where the quantities with subscripts ``mon" and ``quad" represent the contribution from monopole and quadrupole radiation. The quadrupole radiation carries energy and angular momentum, while the monopole radiation carries only energy and no angular momentum. This results in $\mathcal{F}_{\rm quad}=\Omega\mathcal{L}_{\rm quad}$ but $\mathcal{F}_{\rm mon}\neq\Omega\mathcal{L}_{\rm mon}=0$. The monopole radiation vanishes in the spin-aligned case. In the spin-precessing systems, however, the non-zero monopole radiation disrupts the identity $\mathcal{F}=\Omega\mathcal{L}$, resulting in stable zero-eccentricity orbits no longer allowed under radiation reaction. A similar effect also appears in Horndeski gravity, calculated by Ref.\,\cite{Chowdhuri2022}, which also involves monopole scalar radiation. This effect originates from the nature of scalar fields in the scalar-tensor gravity theories, and potentially becomes a criterion for distinguishing scalar-tensor gravity from GR in current or future GW observations.

Taking the right-hand side of Eq.\,(\ref{de-dt}) to zero, one obtains the minimum eccentricity $e_{\rm min}$ of the eventual stable orbit
\begin{equation}
\label{e_min}
e_{\rm min}^2=\frac{625}{155648}\zeta x^2
\left[
\frac{2}{\nu}(\chi_1\sin\kappa_1)(\chi_2\sin\kappa_2)\cos\Delta\psi
-\sum_A\frac{m^2}{m_A^2}(\chi_A\sin\kappa_A)^2\right],
\end{equation}
The $e_{\rm min}^2$ depends only on the on-plane component of the SAM vector, $\chi_A\sin\kappa_A$, which automatically disappears in the spin-aligned case. Assuming a precessing system with eccentricity evolves under radiation reaction in dCS gravity, when the initial value exceeds $e_{\min}$, the eccentricity gradually decreases to $e_{\min}$, arriving at a stable orbit, or completing the merger before that. On the contrary, if the initial value is less than $e_{\min}$, the eccentricity increases as the GW frequency increases. 

\section{\label{sec:Conclusions}Conclusions}
This article works in dCS gravity and investigates the gravitational radiation from BBH systems with orbital eccentricity and spin precession. Following Ref.\,\cite{Chatziioannou2017,Loutrel2022}, we decouple the acceleration equation (\ref{acceleration-equation}), describing the evolution of velocity, and the precession equation (\ref{precession-equation}), describing the evolution of SAM vectors in the MSA framework, in which the precession-related variables are regarded as a set of constants. Starting from the definition of binding energy and OAM at 2PN order with dCS correction, Eq.\,(\ref{acceleration-equation}) is re-expressed as Eqs.\,(\ref{dot-r},\,\ref{dot-psi}), and the QK solution is presented in Eqs.\,(\ref{radial-parameterization},\,\ref{psi-parameterization}), following the QK parameterization provided by Klein \textit{et al} \cite{Klein2010,Klein2018}. The conservative parts of scalar and gravitational waveforms are calculated from the linearized field equation and multipole-moment expansion, with the final expressions listed in Eqs.\,(\ref{scalar-waveform},\,\ref{xi-def}). In GW polarzations, $\xi^{(0)}_{+,\times}$ are expanded in terms of nutation angle $\Theta$, with order of magnitude being $\mathcal{O}(v)$, and the low-frequency phase modulations enter each order of Eq.\,(\ref{xi-0}) through variables $\Upsilon$ and $\alpha$. Although the above expressions are not proportional to the coupling parameter $\zeta$, they encode parity violation due to the dCS correction in the precession equation (\ref{precession-equation}). Furthermore, the energy and OAM losses are shown in Eqs.\,(\ref{energy-flux-res},\,\ref{angular-momentum-flux-res}), and the corresponding secular evolutions of orbital elements are shown in Eqs.\,(\ref{dx-dt},\,\ref{de-dt}). Eq.\,(\ref{de-dt}) presents that the stable zero-eccentricity orbit is not allowed in the precessing BBH systems in dCS gravity, due to the monopole scalar radiation, which provides a potential criterion to distinguish dCS gravity and GR. 

What motivates this work is a set of observational facts that the dCS coupling parameter can not be effectively constrained through GW detection based on circular-orbit waveform templates \cite{Nair2019,Perkins2021,HaitianWang2021}. The improvements in source modeling and template construction can, in principle, enhance the constraint ability of GW observation. This work aims to model the orbital motion and GW radiation from the BBH systems. Once the solution to the precession equation (\ref{precession-equation}), providing the analytical solution to $\kappa_A,\psi_A,\Theta,\Upsilon,\alpha$, is combined with our results shown in Eq.\,(\ref{xi-def}), the complete time-domain waveform is routinely constructed. Therefore, three main topics will be included in our future research. The first one is to extend the results in Ref.\,\cite{Loutrel2022} to the eccentric-orbit case, complete the fully analytical time-domain waveforms, and perform the Fourier transform using the stationary phase approximation or some numerical techniques. The second one is to re-perform Bayesian inference using the existing GW catalog based on the newly proposed template, and a tighter constraint on the dCS coupling parameter is expected to be achieved. The third one is to estimate the ability to test the gravitational parity symmetry of the future GW detectors. In conclusion, this work and a series of planned studies are valuable in obtaining tighter constraints on gravitational parity violation through GW observations.

\acknowledgments
We would like to thank Takahiro Tanaka and Antoine Klein for their helpful discussions and comments. This work is supported by the National Key R\&D Program of China (Grant No. 2022YFC2204602), the National Natural Science Foundation of China (Grant No. 12325301 and 12273035), the Science Research Grants from the China Manned Space Project, the 111 Project for "Observational and Theoretical Research on Dark Matter and Dark Energy" (Grant No. B23042).

\bibliographystyle{JHEP}
\bibliography{biblio.bib}

@book{Will2014,
  title         = {{Gravity}},
  author        = {{E. Poisson and C. M. Will}},
  publisher     = "Cambridge University Press", 
  address       = "Cambridge", 
  year          = "2014",
}

@article{Silva2021,
  title = {{Astrophysical and Theoretical Physics Implications from Multimessenger Neutron Star Observations}},
  author = {{Silva, Hector, O} and others},
  journal = {Phys. Rev. Lett.},
  volume = {126},
  issue = {18},
  pages = {181101},
  numpages = {7},
  year = {2021},
  month = {May},
  publisher = {American Physical Society},
  doi = {10.1103/PhysRevLett.126.181101},
  url = {https://link.aps.org/doi/10.1103/PhysRevLett.126.181101}
}

@article{ChernSimons2003,
  title = {{Chern-Simons modification of general relativity}},
  author = {{Jackiw, R. and Pi, S.-Y.}},
  journal = {Phys. Rev. D},
  volume = {68},
  issue = {10},
  pages = {104012},
  numpages = {10},
  year = {2003},
  month = {Nov},
  publisher = {American Physical Society},
  doi = {10.1103/PhysRevD.68.104012},
  url = {https://link.aps.org/doi/10.1103/PhysRevD.68.104012}
}

@article{Deser1982,
  title = {{Three-Dimensional Massive Gauge Theories}},
  author = {{Deser, S. and Jackiw, R. and Templeton, S.}},
  journal = {Phys. Rev. Lett.},
  volume = {48},
  issue = {15},
  pages = {975--978},
  numpages = {0},
  year = {1982},
  month = {Apr},
  publisher = {American Physical Society},
  doi = {10.1103/PhysRevLett.48.975},
  url = {https://link.aps.org/doi/10.1103/PhysRevLett.48.975}
}

@article{Alexander2009,
title = {{Chern–Simons modified general relativity}},
journal = {Physics Reports},
volume = {480},
number = {1},
pages = {1-55},
year = {2009},
issn = {0370-1573},
doi = {https://doi.org/10.1016/j.physrep.2009.07.002},
author = {{Stephon Alexander and Nicolás Yunes}}}

@article{Crisostomi2018,
  title = {{Beyond Lovelock gravity: Higher derivative metric theories}},
  author = {{Crisostomi, M. and Noui, K. and Charmousis, C. and Langlois, D.}},
  journal = {Phys. Rev. D},
  volume = {97},
  issue = {4},
  pages = {044034},
  numpages = {14},
  year = {2018},
  month = {Feb},
  publisher = {American Physical Society},
  doi = {10.1103/PhysRevD.97.044034},
  url = {https://link.aps.org/doi/10.1103/PhysRevD.97.044034}
}

@article{Lee1956,
  title = {{Question of Parity Conservation in Weak Interactions}},
  author = {{Lee, T. D. and Yang, C. N.}},
  journal = {Phys. Rev.},
  volume = {104},
  issue = {1},
  pages = {254--258},
  numpages = {0},
  year = {1956},
  month = {Oct},
  publisher = {American Physical Society},
  doi = {10.1103/PhysRev.104.254},
  url = {https://link.aps.org/doi/10.1103/PhysRev.104.254}
}

@article{Wu1957,
  title = {{Experimental Test of Parity Conservation in Beta Decay}},
  author = {{Wu, C. S. and Ambler, E. and Hayward, R. W. and Hoppes, D. D. and Hudson, R. P.}},
  journal = {Phys. Rev.},
  volume = {105},
  issue = {4},
  pages = {1413--1415},
  numpages = {0},
  year = {1957},
  month = {Feb},
  publisher = {American Physical Society},
  doi = {10.1103/PhysRev.105.1413},
  url = {https://link.aps.org/doi/10.1103/PhysRev.105.1413}
}

@article{Chowdhuri2022,
  title = {{Study of eccentric binaries in Horndeski gravity}},
  author = {{Chowdhuri, Abhishek and Bhattacharyya, Arpan}},
  journal = {Phys. Rev. D},
  volume = {106},
  issue = {6},
  pages = {064046},
  numpages = {16},
  year = {2022},
  month = {Sep},
  publisher = {American Physical Society},
  doi = {10.1103/PhysRevD.106.064046},
  url = {https://link.aps.org/doi/10.1103/PhysRevD.106.064046}
}

@article{Yagi2012pn,
  title = {{Post-Newtonian, quasicircular binary inspirals in quadratic modified gravity}},
  author = {{Yagi, Kent} and others},
  journal = {Phys. Rev. D},
  volume = {85},
  issue = {6},
  pages = {064022},
  numpages = {26},
  year = {2012},
  doi = {10.1103/PhysRevD.85.064022},
  url = {https://link.aps.org/doi/10.1103/PhysRevD.85.064022}
}

@article{Yagi2012gw,
  title = {{Gravitational Waves from Quasicircular Black-Hole Binaries in Dynamical Chern-Simons Gravity}},
  author = {{Yagi, Kent and Yunes, Nicol\'as and Tanaka, Takahiro}},
  journal = {Phys. Rev. Lett.},
  volume = {109},
  issue = {25},
  pages = {251105},
  numpages = {5},
  year = {2012},
  doi = {10.1103/PhysRevLett.109.251105},
  url = {https://link.aps.org/doi/10.1103/PhysRevLett.109.251105}
}

@article{Yagi2013ns,
  title = {{Isolated and binary neutron stars in dynamical Chern-Simons gravity}},
  author = {{Yagi, Kent} and others},
  journal = {Phys. Rev. D},
  volume = {87},
  issue = {8},
  pages = {084058},
  numpages = {27},
  year = {2013},
  doi = {10.1103/PhysRevD.87.084058},
  url = {https://link.aps.org/doi/10.1103/PhysRevD.87.084058}
}

@article{Yagi2012pn_e,
  title = {{Erratum: Post-Newtonian, quasicircular binary inspirals in quadratic modified gravity [Phys. Rev. D 85, 064022 (2012)]}},
  author = {{Yagi, Kent} and others},
  journal = {Phys. Rev. D},
  volume = {93},
  issue = {2},
  pages = {029902},
  numpages = {7},
  year = {2016},
  doi = {10.1103/PhysRevD.93.029902},
  url = {https://link.aps.org/doi/10.1103/PhysRevD.93.029902}
}

@article{ChangfuShi2022,
  title = {{Testing general relativity with TianQin: The prospect of using the inspiral signals of black hole binaries}},
  author = {{Shi, Changfu and Ji, Mujie and Zhang, Jiandong and Mei, Jianwei}},
  journal = {Phys. Rev. D},
  volume = {108},
  issue = {2},
  pages = {024030},
  numpages = {22},
  year = {2023},
  month = {Jul},
  publisher = {American Physical Society},
  doi = {10.1103/PhysRevD.108.024030},
  url = {https://link.aps.org/doi/10.1103/PhysRevD.108.024030}
}

@article{Perkins2021,
  title = {{Improved gravitational-wave constraints on higher-order curvature theories of gravity}},
  author = {{Perkins, Scott E. and Nair, Remya and Silva, Hector O. and Yunes, Nicol\'as}},
  journal = {Phys. Rev. D},
  volume = {104},
  issue = {2},
  pages = {024060},
  numpages = {18},
  year = {2021},
  month = {Jul},
  publisher = {American Physical Society},
  doi = {10.1103/PhysRevD.104.024060},
  url = {https://link.aps.org/doi/10.1103/PhysRevD.104.024060}
}

@article{Nair2019,
  title = {{Fundamental Physics Implications for Higher-Curvature Theories from Binary Black Hole Signals in the LIGO-Virgo Catalog GWTC-1}},
  author = {{Nair, Remya and Perkins, Scott and Silva, Hector O. and Yunes, Nicol\'as}},
  journal = {Phys. Rev. Lett.},
  volume = {123},
  issue = {19},
  pages = {191101},
  numpages = {6},
  year = {2019},
  month = {Nov},
  publisher = {American Physical Society},
  doi = {10.1103/PhysRevLett.123.191101},
  url = {https://link.aps.org/doi/10.1103/PhysRevLett.123.191101}
}

@article{HaitianWang2021,
  title = {{Tight constraints on Einstein-dilation-Gauss-Bonnet gravity from GW190412 and GW190814}},
  author = {{Wang, Hai-Tian} and others},
  journal = {Phys. Rev. D},
  volume = {104},
  issue = {2},
  pages = {024015},
  numpages = {8},
  year = {2021},
  month = {Jul},
  publisher = {American Physical Society},
  doi = {10.1103/PhysRevD.104.024015},
  url = {https://link.aps.org/doi/10.1103/PhysRevD.104.024015}
}

@article{Grumiller2008BH,
  title = {P{How do black holes spin in Chern-Simons modified gravity?}},
  author = {{Grumiller, Daniel and Yunes, Nicol\'as}},
  journal = {Phys. Rev. D},
  volume = {77},
  issue = {4},
  pages = {044015},
  numpages = {20},
  year = {2008},
  month = {Feb},
  doi = {10.1103/PhysRevD.77.044015},
  url = {https://link.aps.org/doi/10.1103/PhysRevD.77.044015}
}

@article{Yunes2009BH,
  title = {{Dynamical Chern-Simons modified gravity: Spinning black holes in the slow-rotation approximation}},
  author = {{Yunes, Nicol\'as and Pretorius, Frans}},
  journal = {Phys. Rev. D},
  volume = {79},
  issue = {8},
  pages = {084043},
  numpages = {14},
  year = {2009},
  doi = {10.1103/PhysRevD.79.084043},
  url = {https://link.aps.org/doi/10.1103/PhysRevD.79.084043}
}

@article{Yagi2012BH,
  title = {{Slowly rotating black holes in dynamical Chern-Simons gravity: Deformation quadratic in the spin}},
  author = {{Yagi, Kent and Yunes, Nicol\'as and Tanaka, Takahiro}},
  journal = {Phys. Rev. D},
  volume = {86},
  issue = {4},
  pages = {044037},
  numpages = {20},
  year = {2012},
  doi = {10.1103/PhysRevD.86.044037},
  url = {https://link.aps.org/doi/10.1103/PhysRevD.86.044037}
}

@article{Loutrel2018,
  title = {{Spin-precessing black hole binaries in dynamical Chern-Simons gravity}},
  author = {{Loutrel, Nicholas and Tanaka, Takahiro and Yunes, Nicol\'as}},
  journal = {Phys. Rev. D},
  volume = {98},
  issue = {6},
  pages = {064020},
  numpages = {24},
  year = {2018},
  month = {Sep},
  publisher = {American Physical Society},
  doi = {10.1103/PhysRevD.98.064020},
  url = {https://link.aps.org/doi/10.1103/PhysRevD.98.064020}
}

@article{Loutrel2022,
  title = {{Parity violation in spin-precessing binaries: Gravitational waves from the inspiral of black holes in dynamical Chern-Simons gravity}},
  author = {{Loutrel, Nicholas and Yunes, Nicol\'as}},
  journal = {Phys. Rev. D},
  volume = {106},
  issue = {6},
  pages = {064009},
  numpages = {34},
  year = {2022},
  doi = {10.1103/PhysRevD.106.064009},
  url = {https://link.aps.org/doi/10.1103/PhysRevD.106.064009}
}

@article{Blanchet1995,
  title = {{Gravitational waves from inspiralling compact binaries: Energy loss and waveform to second-post-Newtonian order}},
  author = {{Blanchet, Luc and Damour, Thibault and Iyer, Bala R.}},
  journal = {Phys. Rev. D},
  volume = {51},
  issue = {10},
  pages = {5360--5386},
  numpages = {0},
  year = {1995},
  doi = {10.1103/PhysRevD.51.5360},
  url = {https://link.aps.org/doi/10.1103/PhysRevD.51.5360}
}

@article{Faye2006,
  title = {{Higher-order spin effects in the dynamics of compact binaries. I. Equations of motion}},
  author = {{Faye, Guillaume and Blanchet, Luc and Buonanno, Alessandra}},
  journal = {Phys. Rev. D},
  volume = {74},
  issue = {10},
  pages = {104033},
  numpages = {19},
  year = {2006},
  month = {Nov},
  publisher = {American Physical Society},
  doi = {10.1103/PhysRevD.74.104033},
  url = {https://link.aps.org/doi/10.1103/PhysRevD.74.104033}
}

@article{Blanchet2006,
  title = {{Higher-order spin effects in the dynamics of compact binaries. II. Radiation field}},
  author = {{Blanchet, Luc and Buonanno, Alessandra and Faye, Guillaume}},
  journal = {Phys. Rev. D},
  volume = {74},
  issue = {10},
  pages = {104034},
  numpages = {17},
  year = {2006},
  month = {Nov},
  publisher = {American Physical Society},
  doi = {10.1103/PhysRevD.74.104034},
  url = {https://link.aps.org/doi/10.1103/PhysRevD.74.104034}
}

@article{Bohe2013,
	doi = {10.1088/0264-9381/30/13/135009},
	url = {https://doi.org/10.1088/0264-9381/30/13/135009},
	year = 2013,
	volume = {30},
	number = {13},
	pages = {135009},
	author = {{Alejandro Boh{\'{e}} and Sylvain Marsat and Luc Blanchet}},
	title = {{Next-to-next-to-leading order spin{\textendash}orbit effects in the gravitational wave flux and orbital phasing of compact binaries}},
	journal = {Classical and Quantum Gravity}
}

@article{ThorneHartle1985,
  title = {{Laws of motion and precession for black holes and other bodies}},
  author = {{Thorne, Kip S. and Hartle, James B.}},
  journal = {Phys. Rev. D},
  volume = {31},
  issue = {8},
  pages = {1815--1837},
  numpages = {0},
  year = {1985},
  month = {Apr},
  publisher = {American Physical Society},
  doi = {10.1103/PhysRevD.31.1815},
  url = {https://link.aps.org/doi/10.1103/PhysRevD.31.1815}
}

@article{Bohe2015,
	doi = {10.1088/0264-9381/32/19/195010},
	url = {https://doi.org/10.1088/0264-9381/32/19/195010},
	year = 2015,
	volume = {32},
	number = {19},
	pages = {195010},
	author = {{Alejandro Boh{\'{e}} and Guillaume Faye and Sylvain Marsat and Edward K Porter}},
	title = {{Quadratic-in-spin effects in the orbital dynamics and gravitational-wave energy flux of compact binaries at the 3PN order}},
	journal = {Classical and Quantum Gravity}
}

@article{Cho2021,
  title = {{Gravitational radiation from inspiralling compact objects: Spin-spin effects completed at the next-to-leading post-Newtonian order}},
  author = {{Cho, Gihyuk and Pardo, Brian and Porto, Rafael A.}},
  journal = {Phys. Rev. D},
  volume = {104},
  issue = {2},
  pages = {024037},
  numpages = {33},
  year = {2021},
  month = {Jul},
  publisher = {American Physical Society},
  doi = {10.1103/PhysRevD.104.024037},
  url = {https://link.aps.org/doi/10.1103/PhysRevD.104.024037}
}

@misc{Cho2022,
author = {{Cho, Gihyuk and Porto, Rafael A. and Yang, Zixin}},
      journal = {arXiv e-prints},
         year = 2022,
        month = jan,
          eid = {arXiv:2201.05138},
        pages = {arXiv:2201.05138},
archivePrefix = {arXiv},
       eprint = {2201.05138},
}

@article{MingzheLi2020NY,
doi = {10.1088/1475-7516/2020/11/023},
url = {https://dx.doi.org/10.1088/1475-7516/2020/11/023},
year = {2020},
month = {nov},
publisher = {},
volume = {2020},
number = {11},
pages = {023},
author = {{Mingzhe Li and Haomin Rao and Dehao Zhao}},
title = {{A simple parity violating gravity model without ghost instability}},
journal = {Journal of Cosmology and Astroparticle Physics},
}

@article{MingzheLi2021NY,
  title = {{Revisiting a parity violating gravity model without ghost instability: Local Lorentz variance}},
  author = {{Li, Mingzhe and Rao, Haomin and Tong, Yeheng}},
  journal = {Phys. Rev. D},
  volume = {104},
  issue = {8},
  pages = {084077},
  numpages = {13},
  year = {2021},
  month = {Oct},
  publisher = {American Physical Society},
  doi = {10.1103/PhysRevD.104.084077},
  url = {https://link.aps.org/doi/10.1103/PhysRevD.104.084077}
}

@article{Chatzistavrakidis2022,
  title = {{Axion gravitodynamics, Lense-Thirring effect, and gravitational waves}},
  author = {{Chatzistavrakidis, Athanasios and Karagiannis, Georgios and Manolakos, George and Schupp, Peter}},
  journal = {Phys. Rev. D},
  volume = {105},
  issue = {10},
  pages = {104029},
  numpages = {8},
  year = {2022},
  month = {May},
  publisher = {American Physical Society},
  doi = {10.1103/PhysRevD.105.104029},
  url = {https://link.aps.org/doi/10.1103/PhysRevD.105.104029}
}

@article{Bombacigno2023,
doi = {10.1088/1475-7516/2023/02/009},
url = {https://dx.doi.org/10.1088/1475-7516/2023/02/009},
year = {2023},
month = {feb},
publisher = {IOP Publishing},
volume = {2023},
number = {02},
pages = {009},
author = {{F. Bombacigno and F. Moretti and S. Boudet and Gonzalo J. Olmo}},
title = {{Landau damping for gravitational waves in parity-violating theories}},
journal = {Journal of Cosmology and Astroparticle Physics},
}

@article{JinQiao2019,
  title = {{Waveform of gravitational waves in the ghost-free parity-violating gravities}},
  author = {{Qiao, Jin and others}},
  journal = {Phys. Rev. D},
  volume = {100},
  issue = {12},
  pages = {124058},
  numpages = {9},
  year = {2019},
  month = {Dec},
  publisher = {American Physical Society},
  doi = {10.1103/PhysRevD.100.124058},
  url = {https://link.aps.org/doi/10.1103/PhysRevD.100.124058}
}

@ARTICLE{JinQiao2020,
       author = {{Qiao, Jin and Zhu, Tao and Zhao, Wen and Wang, Anzhong}},
        title = "{Polarized primordial gravitational waves in the ghost-free parity-violating gravity}",
      journal = {Phys. Rev. D},
         year = 2020,
        month = feb,
       volume = {101},
       number = {4},
          eid = {043528},
        pages = {043528},
          doi = {10.1103/PhysRevD.101.043528},
archivePrefix = {arXiv},
       eprint = {1911.01580},
 primaryClass = {astro-ph.CO},
       adsurl = {https://ui.adsabs.harvard.edu/abs/2020PhRvD.101d3528Q},
}

@ARTICLE{Takahashi2009,
       author = {{Takahashi, Tomohiro and Soda, Jiro}},
        title = "{Chiral Primordial Gravitational Waves from a Lifshitz Point}",
      journal = {Phys. Rev. Lett},
         year = 2009,
        month = jun,
       volume = {102},
       number = {23},
          eid = {231301},
        pages = {231301},
          doi = {10.1103/PhysRevLett.102.231301},
archivePrefix = {arXiv},
       eprint = {0904.0554},
 primaryClass = {hep-th},
       adsurl = {https://ui.adsabs.harvard.edu/abs/2009PhRvL.102w1301T},
}

@ARTICLE{TaoZhu2013,
       author = {{Zhu, Tao and Zhao, Wen and Huang, Yongqing and Wang, Anzhong and Wu, Qiang}},
        title = "{Effects of parity violation on non-Gaussianity of primordial gravitational waves in Ho{\v{r}}ava-Lifshitz gravity}",
      journal = {Phys. Rev. D},
         year = 2013,
        month = sep,
       volume = {88},
       number = {6},
          eid = {063508},
        pages = {063508},
          doi = {10.1103/PhysRevD.88.063508},
archivePrefix = {arXiv},
       eprint = {1305.0600},
 primaryClass = {hep-th},
       adsurl = {https://ui.adsabs.harvard.edu/abs/2013PhRvD..88f3508Z},
}

@ARTICLE{AnzhongWang2013,
       author = {{Wang, Anzhong and Wu, Qiang and Zhao, Wen and Zhu, Tao}},
        title = "{Polarizing primordial gravitational waves by parity violation}",
      journal = {Phys. Rev. D},
      year = 2013,
        month = may,
       volume = {87},
       number = {10},
          eid = {103512},
        pages = {103512},
          doi = {10.1103/PhysRevD.87.103512},
archivePrefix = {arXiv},
       eprint = {1208.5490},
 primaryClass = {astro-ph.CO},
       adsurl = {https://ui.adsabs.harvard.edu/abs/2013PhRvD..87j3512W}
}

@article{Conroy2019,
doi = {10.1088/1475-7516/2019/12/016},
url = {https://dx.doi.org/10.1088/1475-7516/2019/12/016},
year = {2019},
month = {dec},
publisher = {},
volume = {2019},
number = {12},
pages = {016},
author = {{Aindriú Conroy and Tomi Koivisto}},
title = {{Parity-violating gravity and GW170817 in non-Riemannian cosmology}},
journal = {Journal of Cosmology and Astroparticle Physics},
}

@article{MingzheLi2022,
title = {{A simple parity violating model in the symmetric teleparallel gravity and its cosmological perturbations}},
journal = {Physics Letters B},
volume = {827},
pages = {136968},
year = {2022},
issn = {0370-2693},
doi = {https://doi.org/10.1016/j.physletb.2022.136968},
url = {https://www.sciencedirect.com/science/article/pii/S0370269322001022},
author = {{Mingzhe Li and Dehao Zhao}},
}

@article{MingzheLi2022STG,
  title = {{Possible consistent model of parity violations in the symmetric teleparallel gravity}},
  author = {{Li, Mingzhe and Tong, Yeheng and Zhao, Dehao}},
  journal = {Phys. Rev. D},
  volume = {105},
  issue = {10},
  pages = {104002},
  numpages = {7},
  year = {2022},
  month = {May},
  publisher = {American Physical Society},
  doi = {10.1103/PhysRevD.105.104002},
  url = {https://link.aps.org/doi/10.1103/PhysRevD.105.104002}
}

@article{XianGao2014,
  title = {{Hamiltonian analysis of spatially covariant gravity}},
  author = {{Gao, Xian}},
  journal = {Phys. Rev. D},
  volume = {90},
  issue = {10},
  pages = {104033},
  numpages = {17},
  year = {2014},
  month = {Nov},
  publisher = {American Physical Society},
  doi = {10.1103/PhysRevD.90.104033},
  url = {https://link.aps.org/doi/10.1103/PhysRevD.90.104033}
}

@article{TaoZhu2023,
  title = {{Polarized primordial gravitational waves in spatial covariant gravities}},
  author = {{Zhu, Tao and Zhao, Wen and Wang, Anzhong}},
  journal = {Phys. Rev. D},
  volume = {107},
  issue = {2},
  pages = {024031},
  numpages = {13},
  year = {2023},
  month = {Jan},
  publisher = {American Physical Society},
  doi = {10.1103/PhysRevD.107.024031},
  url = {https://link.aps.org/doi/10.1103/PhysRevD.107.024031}
}

@ARTICLE{JinQiao2023,
AUTHOR={{Qiao, Jin} and others},
TITLE={{Testing parity symmetry of gravity with gravitational waves}},
JOURNAL={Frontiers in Astronomy and Space Sciences},
VOLUME={9}, 
YEAR={2023},      
URL={https://www.frontiersin.org/articles/10.3389/fspas.2022.1109086},       
DOI={10.3389/fspas.2022.1109086},      
ISSN={2296-987X},   
}

@article{WenZhao2020waveform,
  title = {{Waveform of gravitational waves in the general parity-violating gravities}},
  author = {{Zhao, Wen} and others},
  journal = {Phys. Rev. D},
  volume = {101},
  issue = {2},
  pages = {024002},
  numpages = {13},
  year = {2020},
  month = {Jan},
  publisher = {American Physical Society},
  doi = {10.1103/PhysRevD.101.024002},
  url = {https://link.aps.org/doi/10.1103/PhysRevD.101.024002}
}

@misc{Jenks2023,
      title={{Parameterized Parity Violation in Gravitational Wave Propagation}}, 
      author={{Leah Jenks and Lyla Choi and Macarena Lagos and Nicolás Yunes}},
      year={2023},
      eprint={2305.10478},
      archivePrefix={arXiv},
      primaryClass={gr-qc}
}

@article{Maria2022,
  title = {{Constraining gravitational wave amplitude birefringence and Chern-Simons gravity with GWTC-2}},
  author = {{Okounkova, Maria} and others},
  journal = {Phys. Rev. D},
  volume = {106},
  issue = {4},
  pages = {044067},
  numpages = {17},
  year = {2022},
  month = {Aug},
  publisher = {American Physical Society},
  doi = {10.1103/PhysRevD.106.044067},
  url = {https://link.aps.org/doi/10.1103/PhysRevD.106.044067}
}

@article{TaoZhu2022,
    author = "{Zhu, Tao and Zhao, Wen and Wang, Anzhong}",
    title = "{Gravitational wave constraints on spatial covariant gravities}",
    eprint = "2211.04711",
    archivePrefix = "arXiv",
    primaryClass = "gr-qc",
    doi = "10.1103/PhysRevD.107.044051",
    journal = "Phys. Rev. D",
    volume = "107",
    number = "4",
    pages = "044051",
    year = "2023"
}

@article{Yunes2010PVtest,
  title = {{Testing gravitational parity violation with coincident gravitational waves and short gamma-ray bursts}},
  author = {{Yunes, Nicol\'as and O'Shaughnessy, Richard and Owen, Benjamin J. and Alexander, Stephon}},
  journal = {Phys. Rev. D},
  volume = {82},
  issue = {6},
  pages = {064017},
  numpages = {21},
  year = {2010},
  month = {Sep},
  publisher = {American Physical Society},
  doi = {10.1103/PhysRevD.82.064017},
  url = {https://link.aps.org/doi/10.1103/PhysRevD.82.064017}
}

@article{Mirshekari2012,
  title = {{Constraining Lorentz-violating, modified dispersion relations with gravitational waves}},
  author = {{Mirshekari, Saeed and Yunes, Nicol\'as and Will, Clifford M.}},
  journal = {Phys. Rev. D},
  volume = {85},
  issue = {2},
  pages = {024041},
  numpages = {12},
  year = {2012},
  month = {Jan},
  publisher = {American Physical Society},
  doi = {10.1103/PhysRevD.85.024041},
  url = {https://link.aps.org/doi/10.1103/PhysRevD.85.024041}
}

@article{WenZhao2020test,
  title = {{Model-independent test of the parity symmetry of gravity with gravitational waves}},
  author = {{Zhao, W} and others},
  journal = {Eur. Phys. J. C},
  volume = {80},
  pages = {630},
  numpages = {13},
  year = {2020},
  doi = {https://doi.org/10.1140/epjc/s10052-020-8211-4},
}

@article{QiangWu2022,
  title = {{Constraints on the Nieh-Yan modified teleparallel gravity with gravitational waves}},
  author = {{Wu, Qiang} and others},
  journal = {Phys. Rev. D},
  volume = {105},
  issue = {2},
  pages = {024035},
  numpages = {13},
  year = {2022},
  month = {Jan},
  publisher = {American Physical Society},
  doi = {10.1103/PhysRevD.105.024035},
  url = {https://link.aps.org/doi/10.1103/PhysRevD.105.024035}
}

@article{ChengGong2022,
  title = {{Gravitational wave constraints on Lorentz and parity violations in gravity: High-order spatial derivative cases}},
  author = {{Gong, Cheng} and others},
  journal = {Phys. Rev. D},
  volume = {105},
  issue = {4},
  pages = {044034},
  numpages = {13},
  year = {2022},
  month = {Feb},
  publisher = {American Physical Society},
  doi = {10.1103/PhysRevD.105.044034},
  url = {https://link.aps.org/doi/10.1103/PhysRevD.105.044034}
}

@ARTICLE{YifanWang2022,
       author = {{Wang, Yi-Fan and Brown, Stephanie M. and Shao, Lijing and Zhao, Wen}},
        title = "{Tests of gravitational-wave birefringence with the open gravitational-wave catalog}",
      journal = {Phys. Rev. D},
         year = 2022,
        month = oct,
       volume = {106},
       number = {8},
          eid = {084005},
        pages = {084005},
          doi = {10.1103/PhysRevD.106.084005},
archivePrefix = {arXiv},
       eprint = {2109.09718},
 primaryClass = {astro-ph.HE},
       adsurl = {https://ui.adsabs.harvard.edu/abs/2022PhRvD.106h4005W}
}

@ARTICLE{YifanWang2021,
       author = {{Wang, Yi-Fan and Niu, Rui and Zhu, Tao and Zhao, Wen}},
        title = "{Gravitational Wave Implications for the Parity Symmetry of Gravity in the High Energy Region}",
      journal = {The Astrophysical Journal},
         year = 2021,
        month = feb,
       volume = {908},
       number = {1},
          eid = {58},
        pages = {58},
          doi = {10.3847/1538-4357/abd7a6},
archivePrefix = {arXiv},
       eprint = {2002.05668},
 primaryClass = {gr-qc},
       adsurl = {https://ui.adsabs.harvard.edu/abs/2021ApJ...908...58W}
}

@ARTICLE{ZhichaoZhao2022,
       author = {{Zhao, Zhi-Chao and Cao, Zhoujian and Wang, Sai}},
        title = "{Search for the Birefringence of Gravitational Waves with the Third Observing Run of Advanced LIGO-Virgo}",
      journal = {The Astrophysical Journal},
         year = 2022,
        month = may,
       volume = {930},
       number = {2},
          eid = {139},
        pages = {139},
          doi = {10.3847/1538-4357/ac62d3},
archivePrefix = {arXiv},
       eprint = {2201.02813},
 primaryClass = {gr-qc},
       adsurl = {https://ui.adsabs.harvard.edu/abs/2022ApJ...930..139Z}
}

@article{ZhaoLi2023,
doi = {10.1088/1475-7516/2023/04/006},
url = {https://dx.doi.org/10.1088/1475-7516/2023/04/006},
year = {2023},
month = {apr},
publisher = {IOP Publishing},
volume = {2023},
number = {04},
pages = {006},
author = {{Zhao Li and Jin Qiao and Tan Liu and Tao Zhu and Wen Zhao}},
title = {{Gravitational waveform and polarization
from binary black hole inspiral in dynamical Chern-Simons gravity: from generation to propagation}},
journal = {Journal of Cosmology and Astroparticle Physics},
}

@article{ZhaoLi2024,
doi = {10.1088/1475-7516/2024/05/073},
url = {https://dx.doi.org/10.1088/1475-7516/2024/05/073},
year = {2024},
month = {may},
publisher = {IOP Publishing},
volume = {2024},
number = {05},
pages = {073},
author = {{Li, Zhao} and others},
title = {{Gravitational radiation from eccentric binary black hole system in dynamical Chern-Simons gravity}},
journal = {Journal of Cosmology and Astroparticle Physics},
}

@article{Wex1995,
doi = {10.1088/0264-9381/12/4/009},
url = {https://dx.doi.org/10.1088/0264-9381/12/4/009},
year = {1995},
month = {apr},
publisher = {},
volume = {12},
number = {4},
pages = {983},
author = {{Norbert Wex}},
title = {{The second post-Newtonian motion of compact binary-star systems with spin}},
journal = {Classical and Quantum Gravity},
}

@article{Memmesheimer2004,
  title = {{Third post-Newtonian accurate generalized quasi-Keplerian parametrization for compact binaries in eccentric orbits}},
  author = {{Memmesheimer, Raoul-Martin and Gopakumar, Achamveedu and Sch\"afer, Gerhard}},
  journal = {Phys. Rev. D},
  volume = {70},
  issue = {10},
  pages = {104011},
  numpages = {17},
  year = {2004},
  month = {Nov},
  publisher = {American Physical Society},
  doi = {10.1103/PhysRevD.70.104011},
  url = {https://link.aps.org/doi/10.1103/PhysRevD.70.104011}
}

@article{Klein2010,
  title = {{Spin effects in the phasing of gravitational waves from binaries on eccentric orbits}},
  author = {{Klein, Antoine and Jetzer, Philippe}},
  journal = {Phys. Rev. D},
  volume = {81},
  issue = {12},
  pages = {124001},
  numpages = {7},
  year = {2010},
  month = {Jun},
  publisher = {American Physical Society},
  doi = {10.1103/PhysRevD.81.124001},
  url = {https://link.aps.org/doi/10.1103/PhysRevD.81.124001}
}

@article{Chatziioannou2017,
  title = {{Constructing gravitational waves from generic spin-precessing compact binary inspirals}},
  author = {{Chatziioannou, Katerina and Klein, Antoine and Yunes, Nicol\'as and Cornish, Neil}},
  journal = {Phys. Rev. D},
  volume = {95},
  issue = {10},
  pages = {104004},
  numpages = {24},
  year = {2017},
  month = {May},
  publisher = {American Physical Society},
  doi = {10.1103/PhysRevD.95.104004},
  url = {https://link.aps.org/doi/10.1103/PhysRevD.95.104004}
}

@article{Loutrel2019,
doi = {10.1088/1361-6382/aaf2a9},
url = {https://dx.doi.org/10.1088/1361-6382/aaf2a9},
year = {2018},
month = {dec},
publisher = {IOP Publishing},
volume = {36},
number = {2},
pages = {025004},
author = {{Nicholas Loutrel and Samuel Liebersbach and Nicolás Yunes and Neil Cornish}},
title = {{The eccentric behavior of inspiralling compact binaries}},
journal = {Classical and Quantum Gravity},
}

@article{Klein2018,
  title = {{Fourier domain gravitational waveforms for precessing eccentric binaries}},
  author = {{Klein, Antoine and Boetzel, Yannick and Gopakumar, Achamveedu and Jetzer, Philippe and de Vittori, Lorenzo}},
  journal = {Phys. Rev. D},
  volume = {98},
  issue = {10},
  pages = {104043},
  numpages = {24},
  year = {2018},
  month = {Nov},
  publisher = {American Physical Society},
  doi = {10.1103/PhysRevD.98.104043},
  url = {https://link.aps.org/doi/10.1103/PhysRevD.98.104043}
}

@article{Moore2020,
doi = {10.1088/1361-6382/ab8bb6},
url = {https://dx.doi.org/10.1088/1361-6382/ab8bb6},
year = {2020},
month = {jul},
publisher = {IOP Publishing},
volume = {37},
number = {16},
pages = {165006},
author = {{Blake Moore and Nicolás Yunes}},
title = {{Constraining gravity with eccentric gravitational waves: projected upper bounds and model selection}},
journal = {Classical and Quantum Gravity},
}

@article{Buonanno2013,
  title = {{Spin effects on gravitational waves from inspiraling compact binaries at second post-Newtonian order}},
  author = {{Buonanno, Alessandra and Faye, Guillaume and Hinderer, Tanja}},
  journal = {Phys. Rev. D},
  volume = {87},
  issue = {4},
  pages = {044009},
  numpages = {14},
  year = {2013},
  month = {Feb},
  publisher = {American Physical Society},
  doi = {10.1103/PhysRevD.87.044009},
  url = {https://link.aps.org/doi/10.1103/PhysRevD.87.044009}
}

@article{Kidder1995,
  title = {{Coalescing binary systems of compact objects to (post)${}^{5/2}$-Newtonian order. V. Spin effects}},
  author = {{Kidder, Lawrence E.}},
  journal = {Phys. Rev. D},
  volume = {52},
  issue = {2},
  pages = {821--847},
  numpages = {0},
  year = {1995},
  month = {Jul},
  publisher = {American Physical Society},
  doi = {10.1103/PhysRevD.52.821},
  url = {https://link.aps.org/doi/10.1103/PhysRevD.52.821}
}

@article{Konigsdorffer2005,
  title = {{Post-Newtonian accurate parametric solution to the dynamics of spinning compact binaries in eccentric orbits: The leading order spin-orbit interaction}},
  author = {{K\"onigsd\"orffer, Christian and Gopakumar, Achamveedu}},
  journal = {Phys. Rev. D},
  volume = {71},
  issue = {2},
  pages = {024039},
  numpages = {18},
  year = {2005},
  month = {Jan},
  publisher = {American Physical Society},
  doi = {10.1103/PhysRevD.71.024039},
  url = {https://link.aps.org/doi/10.1103/PhysRevD.71.024039}
}

@article{Arun2009,
  title = {{Higher-order spin effects in the amplitude and phase of gravitational waveforms emitted by inspiraling compact binaries: Ready-to-use gravitational waveforms}},
  author = {{Arun, K. G. and Buonanno, Alessandra and Faye, Guillaume and Ochsner, Evan}},
  journal = {Phys. Rev. D},
  volume = {79},
  issue = {10},
  pages = {104023},
  numpages = {42},
  year = {2009},
  month = {May},
  publisher = {American Physical Society},
  doi = {10.1103/PhysRevD.79.104023},
  url = {https://link.aps.org/doi/10.1103/PhysRevD.79.104023}
}

@article{Gergely1999,
  title = {{Spin-spin effects in radiating compact binaries}},
  author = {{Gergely, L\'aszl\'o \'A.}},
  journal = {Phys. Rev. D},
  volume = {61},
  issue = {2},
  pages = {024035},
  numpages = {9},
  year = {1999},
  month = {Dec},
  publisher = {American Physical Society},
  doi = {10.1103/PhysRevD.61.024035},
  url = {https://link.aps.org/doi/10.1103/PhysRevD.61.024035}
}

@article{Gergely2000,
  title = {{Second post-Newtonian radiative evolution of the relative orientations of angular momenta in spinning compact binaries}},
  author = {{Gergely, L\'aszl\'o \'A.}},
  journal = {Phys. Rev. D},
  volume = {62},
  issue = {2},
  pages = {024007},
  numpages = {6},
  year = {2000},
  month = {Jun},
  publisher = {American Physical Society},
  doi = {10.1103/PhysRevD.62.024007},
  url = {https://link.aps.org/doi/10.1103/PhysRevD.62.024007}
}

@article{Gergely2003,
  title = {{Gravitational radiation reaction in compact binary systems: Contribution of the quadrupole-monopole interaction}},
  author = {{Gergely, L\'aszl\'o \'A. and Keresztes, Zolt\'an}},
  journal = {Phys. Rev. D},
  volume = {67},
  issue = {2},
  pages = {024020},
  numpages = {7},
  year = {2003},
  month = {Jan},
  publisher = {American Physical Society},
  doi = {10.1103/PhysRevD.67.024020},
  url = {https://link.aps.org/doi/10.1103/PhysRevD.67.024020}
}

@misc{Adak2006,
      title={The Symmetric Teleparallel Gravity}, 
      author={M. Adak},
      year={2006},
      eprint={gr-qc/0611077},
      archivePrefix={arXiv},
      primaryClass={gr-qc},
      url={https://arxiv.org/abs/gr-qc/0611077}, 
}

@article{Bahamonde2023,
  title={Teleparallel gravity: from theory to cosmology},
  author={Bahamonde, Sebastian and Dialektopoulos, Konstantinos F and Escamilla-Rivera, Celia and Farrugia, Gabriel and Gakis, Viktor and Hendry, Martin and Hohmann, Manuel and Said, Jackson Levi and Mifsud, Jurgen and Di Valentino, Eleonora},
  journal={Reports on Progress in Physics},
  volume={86},
  number={2},
  pages={026901},
  year={2023},
  publisher={IOP Publishing}
}

@article{Maluf2013,
author = {Maluf, José W.},
title = {The teleparallel equivalent of general relativity},
journal = {Annalen der Physik},
volume = {525},
number = {5},
pages = {339-357},
doi = {https://doi.org/10.1002/andp.201200272},
url = {https://onlinelibrary.wiley.com/doi/abs/10.1002/andp.201200272},
eprint = {https://onlinelibrary.wiley.com/doi/pdf/10.1002/andp.201200272},
year = {2013}
}

@misc{Nester1999,
      title={Symmetric teleparallel general relativity}, 
      author={J. M. Nester and H-J Yo},
      year={1999},
      eprint={gr-qc/9809049},
      archivePrefix={arXiv},
      primaryClass={gr-qc},
      url={https://arxiv.org/abs/gr-qc/9809049}, 
}

@misc{Chung2025,
title={Probing quadratic gravity with black-hole ringdown gravitational waves measured by LIGO-Virgo-KAGRA detectors}, 
author={Adrian Ka-Wai Chung and Nicolás Yunes},
year={2025},
eprint={2506.14695},
archivePrefix={arXiv},
primaryClass={gr-qc},
url={https://arxiv.org/abs/2506.14695}, 
}

@article{Silva2023,
  title = {Black-hole ringdown as a probe of higher-curvature gravity theories},
  author = {Silva, Hector O. and Ghosh, Abhirup and Buonanno, Alessandra},
  journal = {Phys. Rev. D},
  volume = {107},
  issue = {4},
  pages = {044030},
  numpages = {22},
  year = {2023},
  month = {Feb},
  publisher = {American Physical Society},
  doi = {10.1103/PhysRevD.107.044030},
  url = {https://link.aps.org/doi/10.1103/PhysRevD.107.044030}
}

@article{Chung2024,
  title = {Spectral method for metric perturbations of black holes: Kerr background case in general relativity},
  author = {Chung, Adrian Ka-Wai and Wagle, Pratik and Yunes, Nicol\'as},
  journal = {Phys. Rev. D},
  volume = {109},
  issue = {4},
  pages = {044072},
  numpages = {29},
  year = {2024},
  month = {Feb},
  publisher = {American Physical Society},
  doi = {10.1103/PhysRevD.109.044072},
  url = {https://link.aps.org/doi/10.1103/PhysRevD.109.044072}
}

@article{Chung2024b,
  title = {Quasinormal mode frequencies and gravitational perturbations of black holes with any subextremal spin in modified gravity through METRICS: The scalar-Gauss-Bonnet gravity case},
  author = {Chung, Adrian Ka-Wai and Yunes, Nicol\'as},
  journal = {Phys. Rev. D},
  volume = {110},
  issue = {6},
  pages = {064019},
  numpages = {33},
  year = {2024},
  month = {Sep},
  publisher = {American Physical Society},
  doi = {10.1103/PhysRevD.110.064019},
  url = {https://link.aps.org/doi/10.1103/PhysRevD.110.064019}
}

@article{Chung2025a,
  title = {Quasinormal mode frequencies and gravitational perturbations of spinning black holes in modified gravity through METRICS: The dynamical Chern-Simons gravity case},
  author = {Chung, Adrian Ka-Wai and Lam, Kelvin Ka-Ho and Yunes, Nicol\'as},
  journal = {Phys. Rev. D},
  volume = {111},
  issue = {12},
  pages = {124052},
  numpages = {23},
  year = {2025},
  month = {Jun},
  publisher = {American Physical Society},
  doi = {10.1103/g83n-rrlj},
  url = {https://link.aps.org/doi/10.1103/g83n-rrlj}
}

\appendix

\section{Complicated expressions involved in gravitational waveforms}

\subsection{\label{Sigma-expression}Waveforms $\Sigma_{+,\times}^{(k)}$ at each PN order in Eq.\,(\ref{xi-0})}
{\small
\begin{subequations}
\begin{align}
&\begin{aligned}
\Sigma^{(0)}_{+}&=\frac{1}{2}\frac{\xi}{1-e_r^2}e_r^2\sin^2\iota\cos V
+\frac{\xi}{1-e_r^2}\Bigg\{-\frac{1}{2}e_r^2\cos2(\beta V+\alpha+\omega+\Upsilon)\\
&-\frac{5}{4}e_r\cos[(1+2\beta)V+2(\alpha+\omega+\Upsilon)]-\cos[2(1+\beta)V+2(\alpha+\omega+\Upsilon)]\\
&-\frac{1}{4}e_r\cos[(3+2\beta)V+2(\alpha+\omega+\Upsilon)]\Bigg\}(1+\cos^2\iota),
\end{aligned}\\
&\begin{aligned}
\Sigma^{(1/2)}_{+}&=\frac{\xi}{1-e_r^2}
\Bigg\{2e_r^2\cos(\beta V+\alpha)\cos(\beta V+\alpha+\omega+\Upsilon)
+e_r\cos(\omega+\Upsilon)\\
&+\frac{5}{2}e_r\cos[(1+2\beta)V+2\alpha+\omega+\Upsilon]
+2\cos[2(1+\beta)V+2\alpha+\omega+\Upsilon]\\
&+\frac{1}{2}e_r\cos[(3+2\beta)V+2\alpha+\omega+\Upsilon]\Bigg\}\sin\iota\cos\iota,
\end{aligned}\\
&\begin{aligned}
\Sigma^{(1)}_{+}&=-\frac{3}{2}\frac{\xi}{1-e_r^2}\Bigg\{e_r^2\cos^2(\beta V+\alpha)
+\frac{1}{2}e_r\cos V
+\frac{5}{4}e_r\cos[(1+2\beta)V+2\alpha]\\
&+\cos[2(1+\beta)V+2\alpha]
+\frac{1}{2}e_r\cos[(3+2\beta)V+2\alpha]\Bigg\}\sin^2\iota\\
&+\frac{\xi}{1-e_r^2}\Bigg\{\frac{1}{2}e_r^2 \cos(\beta V+\alpha)\cos[\beta  V+\alpha+2(\omega+\Upsilon)]
+\frac{1}{4}e_r\cos2(\omega+\Upsilon)\cos V\\
&+\frac{5}{8}e_r\cos[(1+2\beta)V+2(\alpha+\omega+\Upsilon)]
+\frac{1}{2}\cos[2(1+\beta)V+2(\alpha+\omega+\Upsilon)]\\
&+\frac{1}{8}e_r\cos[(3+2\beta)V+2(\alpha+\omega+\Upsilon)]\Bigg\}(1+\cos^2\iota),
\end{aligned}\\
&\begin{aligned}
\Sigma^{(3/2)}_{+}&=\frac{\xi}{1-e_r^2}\left\{-e_r^2\cos^2(\beta V+\alpha)
-\frac{1}{2}e_r\cos V
-\frac{5}{4}e_r\cos[(1+2\beta)V+2\alpha]\right.\\
&\left.-\cos[2(1+\beta)V+2\alpha]
-\frac{1}{4}e_r\cos[(3+2\beta)V+2\alpha]\right\}\sin\iota\cos\iota\cos(\omega+\Upsilon),
\end{aligned}\\
&\begin{aligned}
\Sigma^{(2)}_{+}&=-\frac{1}{16}\frac{\xi}{1-e_r^2}\Bigg\{e_r^2\sin2(\beta V+\alpha)
+\frac{5}{2}e_r\sin[(1+2\beta)V+2\alpha]\\
&+2\sin[2(1+\beta)V+2\alpha]
+\frac{1}{2}e_r\sin[(3+2\beta)V+2\alpha]\Bigg\}(1+\cos^2\iota)\sin2(\omega+\Upsilon),\\
\end{aligned}
\end{align}
\end{subequations}}

{\small
\begin{subequations}
\begin{align}
&\begin{aligned}
\Sigma^{(0)}_{\times}&=\frac{\xi}{1-e_r^2}\Bigg\{-e_r^2\sin2(\beta V+\alpha+\omega+\Upsilon)
-\frac{5}{2}e_r\sin[(1+2\beta)V+2(\alpha+\omega+\Upsilon)]\\
&-2\sin[2(1+\beta)V+2(\alpha+\omega+\Upsilon)]-\frac{1}{2}e_r\sin[(3+2\beta)V+2(\alpha+\omega+\Upsilon)]\Bigg\}\cos\iota,
\end{aligned}\\
&\begin{aligned}
\Sigma^{(1/2)}_{\times}&=\frac{\xi}{1-e_r^2}\Bigg\{2e_r^2\cos(\beta V+\alpha)\sin(\beta V+\alpha+\omega+\Upsilon)\\
&+e_r\cos V\sin(\omega+\Upsilon)
+\frac{5}{2}e_r\sin[(1+2\beta)V+2\alpha+\omega+\Upsilon]\\
&+2\sin[2(1+\beta)V+2\alpha+\omega+\Upsilon]
+\frac{1}{2}e_r\sin[(3+2\beta)V+2\alpha+\omega+\Upsilon]\Bigg\}\sin\iota,
\end{aligned}\\
&\begin{aligned}
\Sigma^{(1)}_{\times}&=\frac{\xi}{1-e_r^2}\Bigg\{e_r^2\cos(\beta V+\alpha)\sin[\beta V+\alpha+2(\omega+\Upsilon)] 
+\frac{1}{2}e_r\cos V\sin2(\omega+\Upsilon)\\
&+\frac{5}{4}e_r\sin[(1+2\beta)V+2(\alpha+\omega+\Upsilon)]
+\sin[2(1+\beta)V+2(\alpha+\omega+\Upsilon)]\\
&+\frac{1}{4}e_r\sin[(3+2\beta)V+2(\alpha+\omega+\Upsilon)]\Bigg\}\cos\iota,
\end{aligned}\\
&\begin{aligned}
\Sigma^{(3/2)}_{\times}&=\frac{\xi}{1-e_r^2}\Bigg\{-e_r^2\cos^2(\beta V+\alpha)+\frac{1}{2}e_r\cos V
-\frac{5}{4}e_r\cos[(1+2\beta)V+2\alpha]\\
&-\cos[2(1+\beta)V+2\alpha]
-\frac{1}{4}e_r\cos[(3+2\beta)V+2\alpha]\Bigg\}\sin\iota\sin(\omega+\Upsilon),
\end{aligned}\\
&\begin{aligned}
\Sigma^{(2)}_{\times}&=\frac{1}{4}\frac{\xi}{1-e_r^2}\Bigg\{\frac{1}{2}e_r^2\sin2(\beta V+\alpha)+\frac{5}{4}e_r\sin[(1+2\beta)V+2\alpha]\\
&+\sin[2(1+\beta)V+2\alpha]+\frac{1}{4}e_r\sin[(3+2\beta)V+2\alpha]\Bigg\}\cos\iota\cos2(\omega+\Upsilon).
\end{aligned}
\end{align}
\end{subequations}}
\subsection{\label{app-SS}The SS-coupling coefficients involved in Eq.\,(\ref{delta-xi-SS})}
{\small
\begin{subequations}
\begin{align}
&\begin{aligned}
\mathcal{A}^{+}_{0}&=-\frac{1}{2}\Big\{\cos\kappa_1\sin\kappa_2\left[2(2+3e_r^2)\sin(\psi_2+\tilde{\alpha})
+3e_r^2\sin(2\beta V-\psi_2+\tilde{\alpha})\right]+(1\leftrightarrow2)\Big\}\sin\iota\cos\iota\\
&-\frac{1}{96}\Big\{2\Big[16(3+5e_r^2)\cos2(\tilde{\alpha}+\bar{\psi})+2e_r^2(21-4e_r^2)\cos2(\beta V+\tilde{\alpha})\cos\Delta\psi\Big](1+\cos^2\iota)\\
&+\Big[32e_r^2\cos2(\beta V-\bar{\psi})
+8(12+11e_r^2+2e_r^4)\cos\Delta\psi\Big]\sin^2\iota\Big\}\sin\kappa_1\sin\kappa_2\\
&+\frac{1}{12}e_r^2\Big\{(3-4e_r^2)\cos2(\beta V+\tilde{\alpha})(1+\cos^2\iota)
-2(7-2e_r^2)\sin^2\iota\Big\}
\cos\kappa_1\cos\kappa_2,
\end{aligned}\\
&\begin{aligned}
\mathcal{A}^{+}_{1}&=-\frac{1}{2}e_r\Big\{\cos\kappa_1\sin\kappa_2\Big[3 (4+e_r^2)\sin(\psi_2+\tilde{\alpha})\\
&+2(3+e_r^2)\sin(2\beta V-\psi_2+\tilde{\alpha})\Big]+(1\leftrightarrow2)\Big\}\sin\iota\cos\iota\\
&+\frac{1}{96}e_r
\Big\{-\Big[(304+66e_r^2) \cos2(\tilde{\alpha}+2\bar{\psi})\\
&\qquad-75e_r^2 \cos2(\tilde{\alpha}+2\beta  V-\bar{\psi})+56\cos(2\tilde{\alpha}+2\beta V)\cos\Delta\psi\Big](1+\cos^2\iota)\\
&-\Big[8(7+3e_r^2) \cos2(\beta V-\bar{\psi})+60(4+e_r^2)\cos\Delta\psi\Big]\sin^2\iota\Big\}\sin\kappa_1\sin\kappa_2\\
&-\frac{1}{12}e_r\Big\{2(11+6e_r^2) \cos2(\tilde{\alpha}+\beta  V)(1+\cos^2\iota)+3(4+e_r^2)\sin^2\iota\Big\}\cos\kappa_1\cos\kappa_2,
\end{aligned}\\
&\begin{aligned}
\mathcal{A}^{+}_{2}&=-\Big\{\cos\kappa_1\sin\kappa_2\left[3e_r^2\sin(\psi_2+\tilde{\alpha})+(2+3e_r^2)\sin(2\beta V-\psi_2+\tilde{\alpha})\right]+(1\leftrightarrow2)\Big\}\sin\iota\cos\iota\\
&+ \frac{1}{24}\Big\{e_r^2\Big[-34 \cos (2 \tilde{\alpha}+2\bar{\psi})+73 \cos (2 \tilde{\alpha}+4 \beta  V-2\bar{\psi})
+6\cos2(\tilde{\alpha}+\beta  V)\cos\Delta\psi\Big](1+\cos^2\iota)\\
&-\Big[8(1+2e_r^2)\cos2(\beta V-\bar{\psi})+30e_r^2\cos\Delta\psi\Big]\sin^2\iota\Big\}\sin\kappa_1\sin\kappa_2\\
&-\frac{1}{2}\Big\{(4+7 e_r^2)\cos2 (\tilde{\alpha}+\beta  V)(1+\cos^2\iota)+e_r^2\sin^2\iota\Big\}\cos\kappa_1\cos\kappa_2,
\end{aligned}\\
&\begin{aligned}
\mathcal{A}^{+}_{3}&=-\frac{1}{4}e_r \Big\{\cos\kappa_1\sin\kappa_2
\left[2e_r^2\sin(\psi_2+\tilde{\alpha})
+3(4+e_r^2)\sin(2\beta V-\psi_2+\tilde{\alpha})\right]+(1\leftrightarrow2)\Big\}\sin\iota\cos\iota\\
&+\frac{1}{96} e_r \Big\{\Big[-22 e_r^2 \cos (2 \tilde{\alpha}+2\bar{\psi})
+17(24+5e_r^2)\cos2(\tilde{\alpha}+2\beta  V-\bar{\psi})\\
&\qquad+2(4-5e_r^2)\cos2(\tilde{\alpha}+\beta  V)\cos\Delta\psi\Big](1+\cos^2\iota)\\
&-\Big[2(28+9e_r^2) \cos2(\beta  V-\bar{\psi})-20e_r^2\cos\Delta\psi\Big]\sin^2\iota\Big\}\sin\kappa_1\sin\kappa_2\\
&-\frac{1}{24} e_r\Big\{(76+13e_r^2)\cos2(\tilde{\alpha}+\beta  V)(1+\cos^2\iota)+2e_r^2\sin^2\iota\Big\}\cos\kappa_1\cos\kappa_2,
\end{aligned}\\
&\begin{aligned}
\mathcal{A}^{+}_{4}&=-\frac{3}{2}e_r^2 \Big\{\cos\kappa_1\sin\kappa_2
\sin(2\beta V-\psi_2+\tilde{\alpha})+(1\leftrightarrow2)\Big\}\sin\iota\cos\iota\\
&+\frac{1}{48}\Big\{\Big[48(2+3e_r^2)\cos 2(\tilde{\alpha}+2\beta V-\bar{\psi})-10e_r^2 \cos2(\tilde{\alpha}+\beta  V)\cos\Delta\psi\Big](1+\cos^2\iota)\\
&-16e_r^2 \cos (2 \beta  V-2\bar{\psi})\sin^2\iota\Big\}\sin\kappa_1\sin\kappa_2
-\frac{13}{12}e_r^2 \cos2 (\tilde{\alpha}+\beta  V)(1+\cos^2\iota)\cos\kappa_1\cos\kappa_2,
\end{aligned}\\
&\begin{aligned}
\mathcal{A}^{+}_{5}&=-\frac{1}{4}e_r^3 \Big\{\cos\kappa_1\sin\kappa_2\sin(2\beta  V-\psi_2+\tilde{\alpha})+(1\leftrightarrow2)\Big\}\sin\iota\cos\iota\\
&+\frac{1}{96}e_r \Big\{\Big[(232+57e_r^2) \cos2(\tilde{\alpha}+2\beta V-\bar{\psi})
-6e_r^2\cos2(\tilde{\alpha}+\beta  V)\cos\Delta\psi\Big](1+\cos^2\iota)\\
&-6e_r^2\cos2(\beta  V-\bar{\psi})\sin^2\iota\Big\}\sin\kappa_1\sin\kappa_2
-\frac{1}{8}e_r^3\cos2(\tilde{\alpha}+\beta  V)(1+\cos^2\iota)\cos\kappa_1\cos\kappa_2,
\end{aligned}\\
&\begin{aligned}
\mathcal{A}^{+}_{6}&=\frac{25}{24}e_r^2 \cos2(\tilde{\alpha}+2\beta  V-\bar{\psi})(1+\cos^2\iota)\sin\kappa_1\sin\kappa_2,
\end{aligned}\\
&\begin{aligned}
\mathcal{A}^{+}_{7}&=\frac{5}{32}e_r^3 \cos2(\tilde{\alpha}+2\beta  V-\bar{\psi})(1+\cos^2\iota)\sin\kappa_1\sin\kappa_2,
\end{aligned}
\end{align}
\end{subequations}
\begin{subequations}
\begin{align}
&\begin{aligned}
\mathcal{B}^{+}_{1}&=-\frac{1}{2}e_r(6+e_r^2) \Big\{\cos\kappa_1\sin\kappa_2\cos(2 \beta  V-\psi_2+\tilde{\alpha})+(1\leftrightarrow2)\Big\}\sin\iota\cos\iota\\
&-\frac{1}{96} e_r \Big\{\Big[-4(4-9e_r^2) \sin2(\tilde{\alpha}+\bar{\psi})+75e_r^2\sin2(\tilde{\alpha}+2\beta V-\bar{\psi})\\
&-4(14-15e_r^2)\sin2(\tilde{\alpha}+\beta  V)\cos\Delta\psi\Big](1+\cos^2\iota)\\
&-4(14+3e_r^2)\sin2(\beta V-\bar{\psi})\sin^2\iota\Big\}\sin\kappa_1\sin\kappa_2\\
&+\frac{1}{12}e_r(22+21e_r^2)
\sin2(\tilde{\alpha}+\beta V)(1+\cos^2\iota)\cos\kappa_1\cos\kappa_2,
\end{aligned}\\
&\begin{aligned}
\mathcal{B}^{+}_{2}&=-(2+3e_r^2)\Big\{\cos\kappa_1\sin\kappa_2\cos(2\beta V-\psi_2+\tilde{\alpha})+(1\leftrightarrow2)\Big\}\sin\iota\cos\iota\\
&+\frac{1}{24}\Big\{-e_r^2\Big[4\sin2(\tilde{\alpha}+\bar{\psi})+73\sin2(\tilde{\alpha}+2\beta V-\bar{\psi})\\
&+6\sin2(\tilde{\alpha}+\beta V)\cos\Delta\psi\Big](1+\cos^2\iota)
+8(1+2e_r^2) \sin (2 \beta  V-2\bar{\psi})\sin^2\iota\Big\}\sin\kappa_1\sin\kappa_2\\
&+\frac{1}{2}(4+7e_r^2)\sin2(\tilde{\alpha}+\beta  V)(1+\cos^2\iota)\cos\kappa_1\cos\kappa_2,
\end{aligned}\\
&\begin{aligned}
\mathcal{B}^{+}_{3}&=-\frac{3}{4}e_r(4+e_r^2) \Big\{\cos\kappa_1\sin\kappa_2\cos(2\beta V-\psi_2+\tilde{\alpha})+(1\leftrightarrow2)\Big\}\sin\iota\cos\iota\\
&-\frac{1}{96}e_r \Big\{\Big[4e_r^2\sin2(\tilde{\alpha}+\bar{\psi})
+17 \left(5 e_r^2+24\right) \sin (2 \tilde{\alpha}+4 \beta  V-2\bar{\psi})\\
&+2(4-5e_r^2)\sin2(\tilde{\alpha}+\beta  V)\cos\Delta\psi\Big](1+\cos^2\iota)\\
&+2(28+9e_r^2)\sin2(\beta  V-\bar{\psi})\sin^2\iota\Big\}\sin\kappa_1\sin\kappa_2\\
&+\frac{1}{24}e_r(76+13e_r^2)\sin2(\tilde{\alpha}+\beta V)(1+\cos^2\iota)\cos\kappa_1\cos\kappa_2,
\end{aligned}\\
&\begin{aligned}
\mathcal{B}^{+}_{4}&=-\frac{3}{2}e_r^2 \Big\{\cos\kappa_1\sin\kappa_2\cos(2\beta V-\psi_2+\tilde{\alpha})+(1\leftrightarrow2)\Big\}\sin\iota\cos\iota\\
&+\frac{1}{48}\Big\{\Big[-48(2+3e_r^2) \sin 2(\tilde{\alpha}+2\beta V-\bar{\psi})+10 e_r^2\sin2(\tilde{\alpha}+\beta  V)\cos\Delta\psi\Big](1+\cos^2\iota)\\
&+16e_r^2\sin2(\beta V-\bar{\psi})\sin^2\iota\Big\}\sin\kappa_1\sin\kappa_2
+\frac{13}{12}e_r^2\sin2(\tilde{\alpha}+\beta  V)(1+\cos^2\iota)\cos\kappa_1\cos\kappa_2,
\end{aligned}\\
&\begin{aligned}
\mathcal{B}^{+}_{5}&=-\frac{1}{4}e_r^3 \Big\{\cos\kappa_1\sin\kappa_2\cos(2\beta V-\psi_2+\tilde{\alpha})+(1\leftrightarrow2)\Big\}\sin\iota\cos\iota\\
&+\frac{1}{96}\Big\{e_r\Big[-(232+57e_r^2)\sin2(\tilde{\alpha}+2\beta  V-\bar{\psi})
+6e_r^2\sin2(\tilde{\alpha}+\beta  V)\cos\Delta\psi\Big](1+\cos^2\iota)\\
&+6e_r^3\sin(2\beta  V-2\bar{\psi})\sin^2\iota\Big\}\sin\kappa_1\sin\kappa_2
+\frac{1}{8}e_r^3\sin2(\tilde{\alpha}+\beta  V)(1+\cos^2\iota)\cos\kappa_1\cos\kappa_2,
\end{aligned}\\
&\begin{aligned}
\mathcal{B}^{+}_{6}&=-\frac{25}{24}e_r^2 \sin2(\tilde{\alpha}+2\beta  V-\bar{\psi})(1+\cos^2\iota)\sin\kappa_1\sin\kappa_2,
\end{aligned}\\
&\begin{aligned}
\mathcal{B}^{+}_{7}&=-\frac{5}{32}e_r^3\sin2(\tilde{\alpha}+2\beta V-\bar{\psi})(1+\cos^2\iota)\sin\kappa_1\sin\kappa_2.
\end{aligned}
\end{align}
\end{subequations}

\begin{subequations}
\begin{align}
&\begin{aligned}
\mathcal{A}^{\times}_{0}&=\frac{1}{2}\Big\{\cos\kappa_1\sin\kappa_2\left[2(2+3e_r^2) \cos(\tilde{\alpha }+\psi _1)+3e_r^2\cos(\tilde{\alpha }+2\beta V-\psi _1)\right]+(1\leftrightarrow2)\Big\}\sin\iota\\
&-\frac{1}{24}\Big\{16(3+5e_r^2)\sin 2(\bar{\psi }+\tilde{\alpha })+2\left(21-4 e_r^2\right) e_r^2 \sin2(\tilde{\alpha }+\beta  V)\cos \Delta\psi\Big\}\cos\iota\sin\kappa_1\sin\kappa_2\\
&+\frac{1}{6}e_r^2(3-4e_r^2)\sin2 (\tilde{\alpha }+\beta  V)\cos\iota\cos\kappa_1\cos\kappa_2,
\end{aligned}\\
&\begin{aligned}
\mathcal{A}^{\times}_{1}&=\frac{1}{2}e_r\Big\{\cos\kappa_1\sin\kappa_2\left[3 (4+e_r^2) \cos(\tilde{\alpha }+\psi _1)+2 (3+e_r^2) \cos(\tilde{\alpha }+2 \beta  V-\psi _1)\right]+(1\leftrightarrow2)\Big\}\sin\iota\\
&-\frac{1}{48}e_r \Big\{\Big[-75 e_r^2 \sin2(\tilde{\alpha }-\bar{\psi }+2\beta  V)+(304+66e_r^2)\sin2(\bar{\psi }+\tilde{\alpha })\\
&+56\sin2(\tilde{\alpha }+\beta  V)\cos \Delta\psi\Big]\Big\}\cos\iota\sin\kappa_1\sin\kappa_2\\
&-\frac{1}{3}e_r(11+6e_r^2) \sin2(\tilde{\alpha }+\beta  V)\cos\iota\cos\kappa_1\cos\kappa_2,
\end{aligned}\\
&\begin{aligned}
\mathcal{A}^{\times}_{2}&=\Big\{\cos\kappa_1\sin\kappa_2\left[3 e_r^2 \cos \left(\tilde{\alpha }+\psi _1\right)+(2+3e_r^2)\cos(\tilde{\alpha }+2 \beta V-\psi_1)\right]+(1\leftrightarrow2)\Big\}\sin\iota\\
&+\frac{1}{12} e_r^2\Big\{-34\sin2 (\bar{\psi }+\tilde{\alpha})+73\sin2(\tilde{\alpha }-\bar{\psi }+2 \beta  V)\\
&+6\sin2(\tilde{\alpha }+\beta  V)\cos\Delta\psi\Big\}\cos\iota\sin\kappa_1\sin\kappa_2\\
&-(4+7e_r^2) \sin2(\tilde{\alpha}+\beta  V)\cos\iota\cos\kappa_1\cos\kappa_2,
\end{aligned}\\
&\begin{aligned}
\mathcal{A}^{\times}_{3}&=\frac{1}{4}e_r\Big\{\cos\kappa_1\sin\kappa_2\left[2e_r^2 \cos(\tilde{\alpha }+\psi _1)+3(4+e_r^2)\cos(\tilde{\alpha }+2 \beta  V-\psi_1)\right]+(1\leftrightarrow2)\Big\}\sin\iota\\
&+\frac{1}{48}e_r\Big\{-22e_r^2\sin2 (\bar{\psi }+\tilde{\alpha })+17(24+5e_r^2)\sin2(\tilde{\alpha }-\bar{\psi }+2 \beta V)\\
&+2(4-5 e_r^2) \cos \Delta\psi \sin \left(2 \left(\tilde{\alpha }+\beta  V\right)\right)\Big\}\cos\iota\sin\kappa_1\sin\kappa_2\\
&-\frac{1}{12}e_r(76+13e_r^2)\sin2(\tilde{\alpha }+\beta V)\cos\iota\cos\kappa_1\cos\kappa_2,
\end{aligned}\\
&\begin{aligned}
\mathcal{A}^{\times}_{4}&=\frac{3}{2}e_r^2\Big\{\cos\kappa_1\sin\kappa_2\cos(\tilde{\alpha }+2 \beta  V-\psi_1)+(1\leftrightarrow2)\Big\}\sin\iota\\
&+\Big\{2(2+3 e_r^2)\sin2(\tilde{\alpha }-\bar{\psi }+2\beta V)-\frac{5}{12}e_r^2 \sin2(\tilde{\alpha }+\beta V)\cos\Delta\psi\Big\}\cos\iota\sin\kappa_1\sin\kappa_2\\
&-\frac{13}{6}e_r^2\sin2(\tilde{\alpha}+\beta V)\cos\iota\cos\kappa_1\cos\kappa_2,
\end{aligned}\\
&\begin{aligned}
\mathcal{A}^{\times}_{5}&=\frac{1}{4}e_r^3\Big\{\cos\kappa_1\sin\kappa_2\cos \left(\tilde{\alpha }+2 \beta  V-\psi _1\right)+(1\leftrightarrow2)\Big\}\sin\iota\\
&+\frac{1}{48} e_r \Big\{(232+57 e_r^2+232)\sin2(\tilde{\alpha }-\bar{\psi }+2\beta  V)\\
&-6e_r^2 \sin2(\tilde{\alpha }+\beta  V)\cos\Delta\psi\Big\}\cos\iota\sin\kappa_1\sin\kappa_2\\
&-\frac{1}{4}e_r^3 \sin2(\tilde{\alpha }+\beta  V)\cos\iota\cos\kappa_1\cos\kappa_2,
\end{aligned}\\
&\begin{aligned}
\mathcal{A}^{\times}_{6}&=\frac{25}{12} e_r^2\sin2(\tilde{\alpha }-\bar{\psi }+2 \beta  V)\cos\iota\sin\kappa_1\sin\kappa_2,
\end{aligned}\\
&\begin{aligned}
\mathcal{A}^{\times}_{7}&=\frac{5}{16}e_r^3\sin2(\tilde{\alpha }-\bar{\psi }+2 \beta  V)\cos\iota\sin\kappa_1\sin\kappa_2,
\end{aligned}
\end{align}
\end{subequations}

\begin{subequations}
\begin{align}
&\begin{aligned}
\mathcal{B}^{\times}_{1}&=-\frac{1}{2}e_r\Big\{\cos\kappa_1\sin\kappa_2(6+e_r^2)\sin(\tilde{\alpha }+2 \beta  V-\psi _1)+(1\leftrightarrow2)\Big\}\sin\iota\\
&-\frac{1}{48}e_r\Big\{4(4-9e_r^2)\cos2(\bar{\psi }+\tilde{\alpha})-75e_r^2 \cos2(\tilde{\alpha }-\bar{\psi }+2 \beta  V)\\
&+4(14-15e_r^2)\cos2(\tilde{\alpha}+\beta  V)\cos\Delta\psi\Big\}
\cos\iota\sin\kappa_1\sin\kappa_2\\
&-\frac{1}{6} e_r(22+21 e_r^2)\sin2 (\tilde{\alpha }+\beta  V)\cos\iota\cos\kappa_1\cos\kappa_2,
\end{aligned}\\
&\begin{aligned}
\mathcal{B}^{\times}_{2}&=-\Big\{\cos\kappa_1\sin\kappa_2(2+3e_r^2) \sin(\tilde{\alpha }+2 \beta  V-\psi _1)
+(1\leftrightarrow2)\Big\}\sin\iota\\
&+\frac{1}{12}e_r^2\Big\{4 \cos2(\bar{\psi }+\tilde{\alpha })+73 \cos2(\tilde{\alpha }-\bar{\psi }+2 \beta  V)\\
&+6\cos2(\tilde{\alpha }+\beta  V)\cos\Delta\psi\Big\}\cos\iota\sin\kappa_1\sin\kappa_2\\
&-(4+7e_r^2)\sin2(\tilde{\alpha }+\beta  V)\cos\iota\cos\kappa_1\cos\kappa_2,
\end{aligned}\\
&\begin{aligned}
\mathcal{B}^{\times}_{3}&=-\frac{3}{4}e_r\Big\{\cos\kappa_1\sin\kappa_2
(4+e_r^2) \sin(\tilde{\alpha }+2 \beta  V-\psi _1)+(1\leftrightarrow2)\Big\}\sin\iota\\
&+\frac{1}{48}e_r\Big\{4e_r^2\cos2(\bar{\psi}+\tilde{\alpha})+17(24+5e_r^2) \cos2 (\tilde{\alpha }-\bar{\psi }+2 \beta  V)\\
&+2 (4-5 e_r^2)\cos2(\tilde{\alpha }+\beta  V)\cos\Delta\psi\Big\}\cos\iota\sin\kappa_1\sin\kappa_2\\
&-\frac{1}{12} e_r(76+13 e_r^2) \sin2 (\tilde{\alpha }+\beta  V)\cos\iota\cos\kappa_1\cos\kappa_2,
\end{aligned}\\
&\begin{aligned}
\mathcal{B}^{\times}_{4}&=-\frac{3}{2}e_r^2\Big\{\cos\kappa_1\sin\kappa_2\sin(\tilde{\alpha }+2 \beta  V-\psi_1)+(1\leftrightarrow2)\Big\}\sin\iota\\
&+\Big\{2(2+3 e_r^2+2)\cos2(\tilde{\alpha }-\bar{\psi }+2\beta  V)-\frac{5}{12}e_r^2 \cos2(\tilde{\alpha }+\beta  V)\cos\Delta\psi\Big\}\cos\iota\sin\kappa_1\sin\kappa_2\\
&-\frac{13}{6}e_r^2 \sin2(\tilde{\alpha }+\beta  V)\cos\iota\cos\kappa_1\cos\kappa_2,
\end{aligned}\\
&\begin{aligned}
\mathcal{B}^{\times}_{5}&=-\frac{1}{4}e_r^3\Big\{\cos\kappa_1\sin\kappa_2\sin \left(\tilde{\alpha }+2 \beta  V-\psi_1\right)+(1\leftrightarrow2)\Big\}\sin\iota\\
&+\frac{1}{48} e_r\Big\{(232+57 e_r^2)\cos2(\tilde{\alpha }-\bar{\psi }+2 \beta  V)-6e_r^2\cos2(\tilde{\alpha }+\beta V)\cos \Delta\psi\Big\}\cos\iota\sin\kappa_1\sin\kappa_2\\
&-\frac{1}{4}e_r^3 \sin2(\tilde{\alpha }+\beta  V)\cos\iota\cos\kappa_1\cos\kappa_2,
\end{aligned}\\
&\begin{aligned}
\mathcal{B}^{\times}_{6}&=\frac{25}{12} e_r^2\cos2(\tilde{\alpha }-\bar{\psi }+2 \beta  V)\cos\iota\sin\kappa_1\sin\kappa_2,
\end{aligned}\\
&\begin{aligned}
\mathcal{B}^{\times}_{7}&= \frac{5}{16} e_r^3\cos2(\tilde{\alpha }-\bar{\psi }+2 \beta  V)\cos\iota\sin\kappa_1\sin\kappa_2.
\end{aligned}
\end{align}
\end{subequations}}

\subsection{\label{app-MQ}The MQ-coupling coefficients involved in Eq.\,(\ref{delta-xi-MQ})}
{\small
\begin{subequations}
\begin{align}
&\begin{aligned}
\mathcal{C}^{+}_{A,0}&=-\frac{1}{16} e_r^2\Big\{(3-4 e_r^2) \cos2 (\tilde{\alpha}+\beta  V)(1+\cos^2\iota)-2(7-2 e_r^2)\sin^2\iota\Big\}\\
&+\frac{1}{32}\Big\{\Big[3e_r^2(15-4e_r^2)\cos2(\tilde{\alpha}+\beta V)+8(3+5e_r^2) \cos2(\tilde{\alpha}+\psi_A)\Big](1+\cos^2\iota)\\
&-\Big[26e_r^2\cos2(\beta  V-\psi_A)
+6(8+5e_r^2+2e_r^4)\Big]\sin^2\iota\Big\}\sin^2\kappa_A\\
&+\frac{3}{8}\Big[2(2+3e_r^2)\sin(\tilde{\alpha}+\psi_A)+3e_r^2\sin (\tilde{\alpha}+2 \beta  V-\psi_A)\Big]\sin\iota\cos\iota\sin\kappa_A\cos\kappa_A,
\end{aligned}\\
&\begin{aligned}
\mathcal{C}^{+}_{A,1}&=\frac{1}{16} e_r \Big\{2(11+6e_r^2)\cos2(\tilde{\alpha}+\beta  V)(1+\cos^2\iota)+3(4+e_r^2)\sin^2\iota\Big\}\\
&+\frac{1}{128}e_r\Big\{\Big[(304+66e_r^2) \cos2(\tilde{\alpha}+\psi_A)-63e_r^2\cos2 (\tilde{\alpha}+2 \beta  V-\psi_A)\\
&+24(1-2 e_r^2)\cos2 (\tilde{\alpha}+\beta  V)\Big](1+\cos^2\iota)\\
&+\Big[432+108 e_r^2+8(25+9e_r^2)\cos (2 \beta  V-2\psi_A)\Big]\sin^2\iota\Big\}\sin^2\kappa_A\\
&+\frac{3}{8}e_r\Big[3(4+e_r^2)\sin(\tilde{\alpha}+\psi_A)+2(3+e_r^2)\sin(\tilde{\alpha}+2\beta  V-\psi_A)\Big]\sin\iota\cos\iota\sin\kappa_A\cos\kappa_A,
\end{aligned}\\
&\begin{aligned}
\mathcal{C}^{+}_{A,2}&=\frac{3}{8} \Big\{(4+7e_r^2) \cos (2 (\tilde{\alpha}+\beta  V))(1+\cos^2\iota)+e_r^2\sin^2\iota\Big\}\\
&+\frac{1}{32}\Big\{\Big[34e_r^2 \cos2 (\tilde{\alpha}+\psi_A)-55e_r^2\cos2 (\tilde{\alpha}+2 \beta  V-\psi_A)\\
&-6(4+9 e_r^2)\cos2(\tilde{\alpha}+\beta  V)\Big](1+\cos^2\iota)\\
&-\Big[54 e_r^2+4(8+13e_r^2)\cos2(\beta V-\psi_A)\Big]\sin^2\iota\Big\}\sin^2\kappa_A\\
&+\frac{3}{4}\Big[3 e_r^2 \sin (\tilde{\alpha}+\psi_A)
+(2+3e_r^2)\sin (\tilde{\alpha}+2 \beta  V-\psi_A)\Big]\sin\iota\cos\iota\sin\kappa_A\cos\kappa_A,
\end{aligned}\\
&\begin{aligned}
\mathcal{C}^{+}_{A,3}&=\frac{1}{32} \Big\{e_r(76+13e_r^2)\cos2(\tilde{\alpha}+\beta  V)(1+\cos^2\iota)+2e_r^3\sin^2\iota\Big\}\\
&+\frac{1}{128}e_r\Big\{\Big[22e_r^2 \cos 2(\tilde{\alpha}+\psi_A)
-49e_r^2\cos2(\tilde{\alpha}+2 \beta  V-\psi_A)
-6(28+e_r^2) \cos2(\tilde{\alpha}+\beta V)\\
&+264\cos2(\tilde{\alpha}+2 \beta  V-\psi_A)\Big](1+\cos^2\iota)\\
&-\Big[36 e_r^2+2(100+27e_r^2)\cos (2 \beta  V-2\psi_A)\Big]\sin^2\iota\Big\}\sin^2\kappa_A\\
&+\frac{3}{16}e_r\Big[2e_r^2 \sin (\tilde{\alpha}+\psi_A)+3(4+e_r^2)\sin(\tilde{\alpha}+2 \beta V-\psi_A)\Big]\sin\iota\cos\iota\sin\kappa_A\cos\kappa_A,
\end{aligned}\\
&\begin{aligned}
\mathcal{C}^{+}_{A,4}&=\frac{13}{16}e_r^2 \cos2(\tilde{\alpha}+\beta  V)(1+\cos^2\iota)\\
&-\frac{1}{32}\Big\{3\Big[4(2+3e_r^2)\cos2 (\tilde{\alpha}+2 \beta  V-\psi_A)
+e_r^2\cos2(\tilde{\alpha}+\beta  V)\Big](1+\cos^2\iota)\\
&-26e_r^2\cos2(\beta V-\psi_A)\sin^2\iota\Big\}\sin^2\kappa_A
+\frac{9}{8}e_r^2\sin(\tilde{\alpha}+2 \beta V-\psi_A)\sin\iota\cos\iota\sin\kappa_A\cos\kappa_A,
\end{aligned}\\
&\begin{aligned}
\mathcal{C}^{+}_{A,5}&=\frac{3}{32}e_r^3 \cos2(\tilde{\alpha}+\beta  V)(1+\cos^2\iota)\\
&+\frac{1}{128} e_r\Big\{\Big[6 e_r^2 \cos2(\tilde{\alpha}+\beta  V)-(88+21e_r^2)\cos2(\tilde{\alpha}+2 \beta  V-\psi_A)\Big](1+\cos^2\iota)\\
&+18 e_r^2 \cos (2 \beta  V-2 \psi_A)\sin^2\iota\Big\}\sin^2\kappa_A
+\frac{3}{16}e_r^3\sin (\tilde{\alpha}+2 \beta  V-\psi_A)\sin\iota\cos\iota\sin\kappa_A\cos\kappa_A,
\end{aligned}\\
&\begin{aligned}
\mathcal{C}^{+}_{A,6}&=-\frac{7}{32}e_r^2\cos2 (\tilde{\alpha}+2 \beta  V-\psi_A)(1+\cos^2\iota)\sin^2\kappa_A,
\end{aligned}\\
&\begin{aligned}
\mathcal{C}^{+}_{A,7}&=-\frac{3}{128}e_r^3\cos2 (\tilde{\alpha}+2 \beta  V-\psi_A)(1+\cos^2\iota)\sin^2\kappa_A,
\end{aligned}
\end{align}
\end{subequations}

\begin{subequations}
\begin{align}
&\begin{aligned}
\mathcal{D}^{+}_{A,1}&=-\frac{1}{16}e_r (22+21 e_r^2)\sin2 (\tilde{\alpha}+\beta  V)(1+\cos^2\iota)\\
&-\frac{1}{128}e_r\Big\{\Big[4(4-9e_r^2)\sin2(\tilde{\alpha}+\psi_A)-63e_r^2\sin2 (\tilde{\alpha}+2 \beta  V-\psi_A)\\
&+12(2-17 e_r^2)\sin2(\tilde{\alpha}+\beta  V)\Big](1+\cos^2\iota)
+4(50+9 e_r^2)\sin2(\beta  V-\psi_A)\sin^2\iota\Big\}\sin^2\kappa_A\\
&+\frac{3}{8}e_r(6+e_r^2)\cos(\tilde{\alpha}+2 \beta  V-\psi_A)\sin\iota\cos\iota\sin\kappa_A\cos\kappa_A,
\end{aligned}\\
&\begin{aligned}
\mathcal{D}^{+}_{A,2}&=-\frac{3}{8}(4+7 e_r^2)\sin2(\tilde{\alpha}+\beta  V)(1+\cos^2\iota)\\
&+\frac{1}{32}\Big\{\Big[4e_r^2\sin2 (\tilde{\alpha}+\psi_A)
+55e_r^2\sin2 (\tilde{\alpha}+2 \beta  V-\psi_A)\\
&+6(4+9 e_r^2)\sin2 (\tilde{\alpha}+\beta  V)\Big](1+\cos^2\iota)
-4 \left(13 e_r^2+8\right)\sin(2 \beta  V-2 \psi_A)\sin^2\iota\Big\}\sin^2\kappa_A\\
&+\frac{3}{4}(2+3 e_r^2)\cos(\tilde{\alpha}+2 \beta  V-\psi_A)\sin\iota\cos\iota\sin\kappa_A\cos\kappa_A,
\end{aligned}\\
&\begin{aligned}
\mathcal{D}^{+}_{A,3}&=-\frac{1}{32}e_r (76+13 e_r^2) \sin2(\tilde{\alpha}+\beta  V)(1+\cos^2\iota)\\
&+\frac{1}{128}e_r\Big\{\Big[4 e_r^2 \sin (2 (\tilde{\alpha}+\psi_A))+(264+49 e_r^2) \sin2 (\tilde{\alpha}+2 \beta  V-\psi_A)\\
&+6(28+e_r^2) \sin2 (\tilde{\alpha}+\beta  V)\Big](1+\cos^2\iota)
-2(100+27e_r^2)\sin2(\beta V-\psi_A)\sin^2\iota\Big\}\sin^2\kappa_A\\
&+\frac{9}{16}e_r(4+e_r^2)\cos(\tilde{\alpha}+2 \beta V-\psi_A)\sin\iota\cos\iota\sin\kappa_A\cos\kappa_A,
\end{aligned}\\
&\begin{aligned}
\mathcal{D}^{+}_{A,4}&=-\frac{13}{16}e_r^2\sin2(\tilde{\alpha}+\beta  V)(1+\cos^2\iota)\\
&+\frac{1}{32}\Big\{+3\Big[4(2+3e_r^2)\sin2 (\tilde{\alpha}+2 \beta  V-\psi_A)+e_r^2\sin2(\tilde{\alpha}+\beta  V)\Big](1+\cos^2\iota)\sin^2\kappa_A\\
&-26e_r^2\sin2(\beta V-\psi_A)\sin^2\iota\Big\}\cos^2\kappa_A
+\frac{9}{8}e_r^2\cos (\tilde{\alpha}+2 \beta  V-\psi_A)\sin\iota\cos\iota\sin\kappa_A\cos\kappa_A,
\end{aligned}\\
&\begin{aligned}
\mathcal{D}^{+}_{A,5}&=-\frac{3}{32}e_r^3 \sin2 (\tilde{\alpha}+\beta  V)(1+\cos^2\iota)\\
&-\frac{1}{128}e_r\Big\{\Big[6 e_r^2 \sin2 (\tilde{\alpha}+\beta V)-(88+21e_r^2)\sin2(\tilde{\alpha}+2 \beta  V-\psi_A)\Big](1+\cos^2\iota)\\
&+18 e_r^2\sin2(\beta V-\psi_A)\sin^2\iota\Big\}\sin^2\kappa_A
+\frac{3}{16}e_r^3\cos(\tilde{\alpha}+2 \beta V-\psi_A)\sin\iota\cos\iota\sin\kappa_A\cos\kappa_A,
\end{aligned}\\
&\begin{aligned}
\mathcal{D}^{+}_{A,6}=\frac{7}{32} e_r^2 \sin2 (\tilde{\alpha}+2 \beta  V-\psi_A)(1+\cos^2\iota)\sin^2\kappa_A,
\end{aligned}\\
&\begin{aligned}
\mathcal{D}^{+}_{A,7}=\frac{3}{128}e_r^3\sin2(\tilde{\alpha}+2 \beta  V-\psi_A)(1+\cos^2\iota)\sin^2\kappa_A,
\end{aligned}
\end{align}
\end{subequations}

\begin{subequations}
\begin{align}
&\begin{aligned}
\mathcal{C}^{\times}_{A,0}&=-\frac{1}{8}e_r^2(3-4e_r^2)\sin2(\tilde{\alpha}+\beta  V)\cos\iota\\
&+\frac{1}{16}\Big\{3e_r^2(15-4e_r^2)\sin2 (\tilde{\alpha}+\beta  V)+8(3+5e_r^2)\sin2 (\tilde{\alpha}+\psi_A)\Big\}\cos\iota\sin^2\kappa_A\\
&-\frac{3}{8}\Big\{2(2+3e_r^2)\cos(\tilde{\alpha}+\psi_A)+3e_r^2\cos(\tilde{\alpha}+2\beta V-\psi_A)\Big\}\sin\iota\sin\kappa_A\cos\kappa_A,
\end{aligned}\\
&\begin{aligned}
\mathcal{C}^{\times}_{A,1}&=\frac{1}{4} e_r(11+6 e_r^2)\sin2(\tilde{\alpha}+\beta  V)\cos\iota\\
&-\frac{1}{64}e_r\Big\{-2(152+33e_r^2)\sin2(\tilde{\alpha}+\psi_A)+63e_r^2\sin2(\tilde{\alpha}+2 \beta  V-\psi_A)\\
&-24(1-2 e_r^2)\sin2(\tilde{\alpha}+\beta  V)\Big\}\cos\iota\sin^2\kappa_A\\
&-\frac{3}{8}e_r\Big\{3(4+e_r^2)\cos (\tilde{\alpha}+\psi_A)+2(3+e_r^2)\cos (\tilde{\alpha}+2 \beta  V-\psi_A)\Big\}\sin\iota\sin\kappa_A\cos\kappa_A,
\end{aligned}\\
&\begin{aligned}
\mathcal{C}^{\times}_{A,2}&=\frac{3}{4}(4+7e_r^2)\sin2(\tilde{\alpha}+\beta  V)\cos\iota\\
&-\frac{1}{16}\Big\{(55e_r^2\sin2(\tilde{\alpha}+2 \beta  V-\psi_A)\\
&-34e_r^2\sin2 (\tilde{\alpha}+\psi_A)+6(4+9 e_r^2)\sin2 (\tilde{\alpha}+\beta  V)\Big\}\cos\iota\sin^2\kappa_A\\
&-\frac{3}{4}\Big\{3 e_r^2 \cos (\tilde{\alpha}+\psi_A)+\left(3 e_r^2+2\right) \cos (\tilde{\alpha}+2 \beta  V-\psi_A)\Big\}\sin\iota\sin\kappa_A\cos\kappa_A,
\end{aligned}\\
&\begin{aligned}
\mathcal{C}^{\times}_{A,3}&=\frac{1}{16}e_r(76+13e_r^2)\sin2(\tilde{\alpha}+\beta  V)\cos\iota\\
&+\frac{1}{64}e_r\Big\{22e_r^2\sin2 (\tilde{\alpha}+\psi_A)-(264+49e_r^2)\sin2 (\tilde{\alpha}+2 \beta  V-\psi_A)\\
&-6(28+e_r^2)\sin2(\tilde{\alpha}+\beta  V)\Big\}\cos\iota\sin^2\kappa_A\\
&-\frac{3}{16}e_r\Big\{2e_r^2\cos (\tilde{\alpha}+\psi_A)+3(4+e_r^2)\cos (\tilde{\alpha}+2 \beta  V-\psi_A)\Big\}\sin\iota\sin\kappa_A\cos\kappa_A,
\end{aligned}\\
&\begin{aligned}
\mathcal{C}^{\times}_{A,4}&=\frac{13}{8}e_r^2 \sin2 (\tilde{\alpha}+\beta V)\cos\iota
-\frac{3}{16}\Big\{4(2+3e_r^2)\sin2 (\tilde{\alpha}+2 \beta  V-\psi_A)\\
&+e_r^2\sin2 (\tilde{\alpha}+\beta  V)\Big\}\cos\iota\sin^2\kappa_A
-\frac{9}{8}e_r^2\cos (\tilde{\alpha}+2 \beta  V-\psi_A)\sin\iota\sin\kappa_A\cos\kappa_A,
\end{aligned}\\
&\begin{aligned}
\mathcal{C}^{\times}_{A,5}&=\frac{3}{16}e_r^3 \sin2(\tilde{\alpha}+\beta  V)\cos\iota
-\frac{1}{64}e_r\Big\{(88+21e_r^2)\sin2 (\tilde{\alpha}+2 \beta  V-\psi_A)\\
&-6e_r^2\sin2(\tilde{\alpha}+\beta  V)\Big\}\cos\iota\sin^2\kappa_A
-\frac{3}{16}e_r^3\cos (\tilde{\alpha}+2 \beta  V-\psi_A)\Big\}\sin\iota\sin\kappa_A\cos\kappa_A,
\end{aligned}\\
&\begin{aligned}
\mathcal{C}^{\times}_{A,6}&=-\frac{7}{16}e_r^2 \sin (2 (\tilde{\alpha}+2 \beta  V-\psi_A))\cos\iota\sin^2\kappa_A,
\end{aligned}\\
&\begin{aligned}
\mathcal{C}^{\times}_{A,7}&=-\frac{3}{64}e_r^3 \sin (2 (\tilde{\alpha}+2 \beta  V-\psi_A))\cos\iota\sin^2\kappa_A,
\end{aligned}
\end{align}
\end{subequations}

\begin{subequations}
\begin{align}
&\begin{aligned}
\mathcal{D}^{\times}_{A,1}&=\frac{1}{8}e_r(22+21e_r^2)\cos (2 (\tilde{\alpha}+\beta  V))\cos\iota
-\frac{1}{64}e_r\Big\{4(4-9 e_r^2)\cos2 (\tilde{\alpha}+\psi_A)\\
&+63e_r^2 \cos2 (\tilde{\alpha}+2 \beta  V-\psi_A)
-12(2-17 e_r^2)\cos2 (\tilde{\alpha}+\beta  V)\Big\}\cos\iota\sin^2\kappa_A\\
&+\frac{3}{8}e_r(6+e_r^2)\sin (\tilde{\alpha}+2 \beta  V-\psi_A)\sin\iota\sin\kappa_A\cos\kappa_A,
\end{aligned}\\
&\begin{aligned}
\mathcal{D}^{\times}_{A,2}&=\frac{3}{4} (4+7 e_r^2) \cos2(\tilde{\alpha}+\beta  V)\cos\iota\\
&-\frac{1}{16}\Big\{4e_r^2\cos2 (\tilde{\alpha}+\psi_A)+55e_r^2\cos2 (\tilde{\alpha}+2 \beta  V-\psi_A)
+6(4+9 e_r^2)\cos2(\tilde{\alpha}+\beta  V)\Big\}\cos\iota\sin^2\kappa_A\\
&+\frac{3}{4}(2+3e_r^2)\sin (\tilde{\alpha}+2\beta  V-\psi_A)\sin\iota\sin\kappa_A\cos\kappa_A,
\end{aligned}\\
&\begin{aligned}
\mathcal{D}^{\times}_{A,3}&=\frac{1}{16}e_r(76+13e_r^2)\cos2(\tilde{\alpha}+\beta  V)\cos\iota
-\frac{1}{64}e_r\Big\{4e_r^2 \cos2(\tilde{\alpha}+\psi_A)\\
&+(264+49e_r^2)\cos2(\tilde{\alpha}+2 \beta  V-\psi_A)
+6(28+e_r^2)\cos2 (\tilde{\alpha}+\beta  V)\Big\}\cos\iota\sin^2\kappa_A\\
&+\frac{9}{16}e_r(4+e_r^2)\sin(\tilde{\alpha}+2\beta V-\psi_A)\sin\iota\sin\kappa_A\cos\kappa_A,
\end{aligned}\\
&\begin{aligned}
\mathcal{D}^{\times}_{A,4}&=\frac{13}{8}e_r^2\cos2 (\tilde{\alpha}+\beta  V)\cos\iota
-\frac{3}{16}\Big\{4(2+3 e_r^2)\cos2 (\tilde{\alpha}+2 \beta  V-\psi_A)\\
&+e_r^2\cos2 (\tilde{\alpha}+\beta  V)\Big\}\cos\iota\sin^2\kappa_A
+\frac{9}{8}e_r^2\sin (\tilde{\alpha}+2 \beta  V-\psi_A)\sin\iota\sin\kappa_A\cos\kappa_A,
\end{aligned}\\
&\begin{aligned}
\mathcal{D}^{\times}_{A,5}&=\frac{3}{16} e_r^3 \cos2 (\tilde{\alpha}+\beta  V)\cos\iota
-\frac{1}{64}e_r\Big\{(88+21e_r^2)\cos2 (\tilde{\alpha}+2 \beta  V-\psi_A)\\
&-6e_r^2\cos2 (\tilde{\alpha}+\beta  V)\Big\}\cos\iota\sin^2\kappa_A
+\frac{3}{16}e_r^3\sin (\tilde{\alpha}+2 \beta V-\psi_A)\sin\iota\sin\kappa_A\cos\kappa_A,
\end{aligned}\\
&\begin{aligned}
\mathcal{D}^{\times}_{A,6}=-\frac{7}{16}e_r^2\cos2(\tilde{\alpha}+2 \beta  V-\psi_A)\cos\iota\sin^2\kappa_A,
\end{aligned}\\
&\begin{aligned}
\mathcal{D}^{\times}_{A,7}=-\frac{3}{64}e_r^3\cos2 (\tilde{\alpha}+2 \beta  V-\psi_A)\cos\iota\sin^2\kappa_A.
\end{aligned}
\end{align}
\end{subequations}}

\end{document}